\documentclass[a4paper,oneside,11pt]{article}
\usepackage{graphicx,afterpage}
\usepackage{bm}\usepackage{soul}
\usepackage{empheq}\usepackage{mathtools}
\usepackage[top=0.6in, bottom=1in, left=1in, right=0.8in]{geometry}
\usepackage{mathrsfs,color}
\usepackage{amsfonts}\usepackage{multirow}
\usepackage{amssymb}\usepackage{caption,comment}
\usepackage{amsmath}\usepackage{float}
\usepackage{amssymb,amsthm}
\usepackage{afterpage,cancel}
\usepackage{tikz,pifont}			% allows us to import images

\usepackage{tikz}
\usepackage[pdftex,colorlinks=true]{hyperref}
\hypersetup{CJKbookmarks,%
bookmarksnumbered,%
colorlinks,%
linkcolor=black,%
citecolor=black,%
plainpages=false,%
pdfstartview=FitH}
\numberwithin{equation}{section}
\numberwithin{figure}{section}

\newcommand\tabcaption{\def\@captype{table}\caption}

\newtheorem{thm}{Theorem}[section]

\newtheorem{prop}[thm]{Proposition}

\definecolor{orange}{RGB}{255,127,0}

\def\d{{\, \rm d}}

\afterpage{\clearpage}
\title{A Mathematical Framework for Quantifying Nonlinear Uncertainty Propagation in Eddy Identification Criteria}
\author{Charlotte Moser, Nan Chen, and Stephen Wiggins}
\date{\today}
\begin{document}
\maketitle %\tableofcontents

\begin{abstract}
Ocean eddies are swirling mesoscale features that play a fundamental role in oceanic transport and mixing. Eddy identification relies on diagnostic criteria that are inherently nonlinear functions of the flow variables. However, estimating the ocean flow field is subject to uncertainty due to its turbulent nature and the use of sparse and noisy observations. This uncertainty interacts with nonlinear diagnostics, complicating its quantification and limiting the accuracy of eddy identification.
In this paper, an analytically tractable mathematical and computational framework for studying eddy identification is developed. It aims to address how uncertainty interacts with the nonlinearity in the eddy diagnostics and how the uncertainty in the eddy diagnostics is reduced when additional information from observations is incorporated.
The framework employs a simple stochastic model for the flow field that mimics turbulent dynamics, allowing closed-form solutions for assessing uncertainty in eddy statistics. It also leverages a nonlinear, yet analytically tractable, data assimilation scheme to incorporate observations, facilitating the study of uncertainty reduction in eddy identification, which is quantified rigorously using information theory.
Applied to the Okubo-Weiss (OW) parameter, a widely used eddy diagnostic criterion, the framework leads to three key results. First, closed formulae reveal inhomogeneous spatial patterns in the OW uncertainty despite homogeneous flow field uncertainty. Second, it shows a close link between local minima of the OW expectation (eddy centers) and local maxima of its uncertainty. Third, it reveals a practical information barrier: the reduction in uncertainty in diagnostics asymptotically saturates, limiting the benefit of additional observations.
\end{abstract}

\section{Introduction}
Oceanic eddies are rotating flow structures generated by turbulent dynamics that propagate across the ocean over a wide range of spatial and temporal scales. The evolution and migration of mesoscale eddies, in particular, are highly influential on ocean dynamics \cite{morrow2012recent, robinson2012eddies, dong2025oceanic}. They drive the transport of momentum, heat, salinity, and mass \cite{zhang2014oceanic, chelton2007global, cheng2014statistical, yankovsky2022influences, koszalka2010vertical, hausmann2012observed}, as well as biogeochemical properties including nutrients \cite{mathis2007eddy, nagai2015dominant, martin2001mechanisms, whitney2002structure, whitney2005physical, woodward2001nutrient, hayward1987nutrient, lewis1986vertical, falkowski1991role, martin2003estimates, siegel1999mesoscale}, oxygen \cite{sarma2018ventilation, eddebbar2021seasonal, atkins2022quantifying, czeschel2018transport}, and biomass \cite{sandulescu2007plankton, salihoglu1990transport, angel1983eddies, prairie2012biophysical, gaube2013satellite, mcgillicuddy1998influence}, in addition to pollutants \cite{viikmae2013impact}. As core components of the ocean system, eddy identification and tracking are therefore crucial for understanding the air-sea interactions that guide global climate \cite{seo2023ocean, dong2025oceanic}.

Eddies are detected and tracked using both observational data and numerical models. Satellite data provides the primary resource for real-time eddy tracking, with sea surface height (SSH) from altimetry being one of the most commonly utilized metrics \cite{zhang2014oceanic, dong2025oceanic, fu2010eddy, le2001ocean, pegliasco2021new, yang2022satellite, chelton2007global, ji2017oceanic}. In regions where SSH is unavailable, such as the Arctic, sea ice floe trajectories extracted from satellite images provide a useful alternative for tracking eddies \cite{lopez2019ice, lopez2021library, manucharyan2022spinning, covington2022bridging}. These satellite datasets are further supplemented by in-situ measurements. Despite being sparse, the in-situ measurements are valuable resources for validating and refining the satellite-derived data \cite{timmermans2008eddies, zhao2016evolution, zhang2015dynamical, subirade2023combining, sandalyuk20203}. Complementing these observations, the dynamics and statistics of eddies are also widely studied using numerical models, such as eddy-resolving general circulation models (GCMs) \cite{zhang2024mesoscale, feng20233d, metzger2009predicting, trott2023luzon, zhang2022physical, shee2024three, rezende2011mean, kang2013gulf} and quasi-geostrophic (QG) models \cite{chen2016eddy, shevchenko2015multi, kim2024characterization}.

There are several eddy identification diagnostics, and the choice of criteria depends on the application of interest. Methods fall into two key classes based on the input data framework: Eulerian and Lagrangian. Eulerian diagnostics use observations at fixed locations, making them directly applicable to satellite altimetry data. The most commonly used Eulerian diagnostic criterion is the Okubo-Weiss (OW) parameter, which identifies eddies using kinematic properties of the flow field \cite{okubo1970horizontal, weiss1991dynamics}. Other widely used Eulerian approaches include those based on the geometry of the velocity vectors \cite{nencioli2010vector}, those that identify closed SSH contours \cite{chelton2011global, faghmous2015daily}, and wavelet analysis of altimetric SSH measurements. In contrast, Lagrangian eddy detection methods incorporate dynamical information from particle trajectories rather than static snapshots, allowing them to capture the evolving coherent structure of eddies \cite{branicki2011lagrangian}. Commonly used Lagrangian techniques include Lagrangian descriptors \cite{mancho2013lagrangian, vortmeyer2016detecting, vortmeyer2019comparing} and finite-time Lyapunov exponents \cite{nolan2020finite}.

A major issue in eddy identification is the inherent uncertainty in observations and model simulations, which degrades the accuracy of the flow field estimate and subsequently limits eddy diagnostic accuracy. Sources of uncertainty in satellite observations include low spatiotemporal resolution and altimetric noise \cite{larnicol1995mean, chelton2011global, fu2010eddy, amores2018up}. The sparse ice floe trajectories in the Arctic also introduce significant uncertainty when used to recover the underlying flow field \cite{lopez2019ice, lopez2021library, manucharyan2022spinning}. On the other hand, models inevitably contain uncertainty originating from assumptions in model structure, parameterizations, and initial conditions, which turbulent dynamics are particularly sensitive to \cite{chen2025taming, majda2018model, majda2012lessons, palmer2001nonlinear, orrell2001model, mignolet2008stochastic, hu2010ensemble, majda2012challenges}. To mitigate these uncertainties, observations and model simulations can be optimally combined using data assimilation \cite{chen2025taming, majda2018model, kalnay2003atmospheric, lahoz2010data, majda2012filtering, stuart2015data, chen2023stochastic, azouani2014continuous}. However, a fundamental challenge remains in understanding the uncertainty in the eddy diagnostic itself. Since eddy identification criteria are inherently nonlinear functions of the flow variables, the uncertainty in the flow field does not directly inform the uncertainty in the eddy diagnostic. This nonlinearity can suppress or amplify uncertainty in unpredictable ways; thus, a Gaussian flow field does not guarantee a Gaussian eddy diagnostic \cite{chen2025taming, resseguier2017geophysical1, resseguier2017geophysical2, resseguier2017geophysical3}.

Recent studies have begun to explore uncertainties in eddy identification. Mainstream methods use ensembles of flow field realizations to visualize uncertainty in eddy diagnostics \cite{raith2021uncertainty, hollt2014ovis, potter2009ensemble, potter2011quantification, chen2024lagrangian}. These studies have highlighted a key issue: eddies identified from the mean flow field can be statistically biased due to the nonlinear propagation of uncertainty \cite{covington2025probabilistic}. However, the ensemble methods are often computationally intractable for high-dimensional systems and cannot always confidently provide systematic guidance. Theoretical studies are therefore essential to overcome these limitations. By developing a rigorous mathematical framework, we can gain a fundamental understanding of how uncertainty propagates through nonlinear eddy identification criteria, which can also advance the development of efficient numerical algorithms that exploit analytically tractable properties.

In this paper, an analytically tractable mathematical framework for studying eddy identification is developed. It aims to address uncertainty in eddy identification through two key perspectives: i) how uncertainty interacts with the nonlinearity in the eddy diagnostic, and ii) how the uncertainty in the eddy diagnostics is reduced when additional information from observations is incorporated. The framework employs a simple stochastic model for the flow field that mimics turbulent dynamics, allowing closed-form solutions for assessing uncertainty in eddy statistics. It also leverages a nonlinear, yet analytically tractable, Lagrangian data assimilation scheme that facilitates the study of uncertainty reduction in eddy identification from assimilating observations, where information theory is incorporated to rigorously quantify such a reduction in uncertainty. The study aims to highlight how the nonlinear nature of these criteria transforms the uncertainty structure in ways that differ markedly from the underlying flow field. Notably, the theoretical foundations built here also advance the development of highly efficient numerical algorithms that bypass computationally prohibitive Monte Carlo approaches for quantifying uncertainty propagation through nonlinear eddy criteria.

The remainder of the paper is structured as follows. Section \ref{sec:framework} introduces the general mathematical framework for eddy identification, including the details of the stochastic flow field model, the Lagrangian data assimilation framework for incorporating observational information into the model simulation, and the information theory approach to quantifying uncertainty reduction. The eddy identification criterion, namely, the OW parameter, used in this work, and the dynamical regimes are introduced in Section \ref{sec:defs}. In Sections \ref{sec:Eddy_UQ}--\ref{sec:spatial_patterns}, the framework is applied to the OW parameter to demonstrate its use. Specifically, in Section \ref{sec:Eddy_UQ}, the framework is utilized to explicitly relate the flow field statistics to the OW parameter statistics. In Section \ref{sec:uncertain_reduce_tracers}, the framework is applied to study the uncertainty in both the flow field and the OW parameter as additional observations are introduced. Section \ref{sec:spatial_patterns} explores the spatial patterns of the OW parameter mean and variance, connecting their fluctuations to the local minimums of the OW parameter. Finally, the paper is concluded in Section \ref{sec:conclusion}.

\section{An Analytically Tractable Mathematical Framework for Eddy Identification}\label{sec:framework}

In this section, we outline the analytically tractable mathematical framework for eddy identification, detailing each tool utilized to examine the uncertainty propagation.
\subsection{Overview of the framework}\label{sec:overview}

\begin{figure}[h!]
    \centering
    \includegraphics[width=\linewidth]{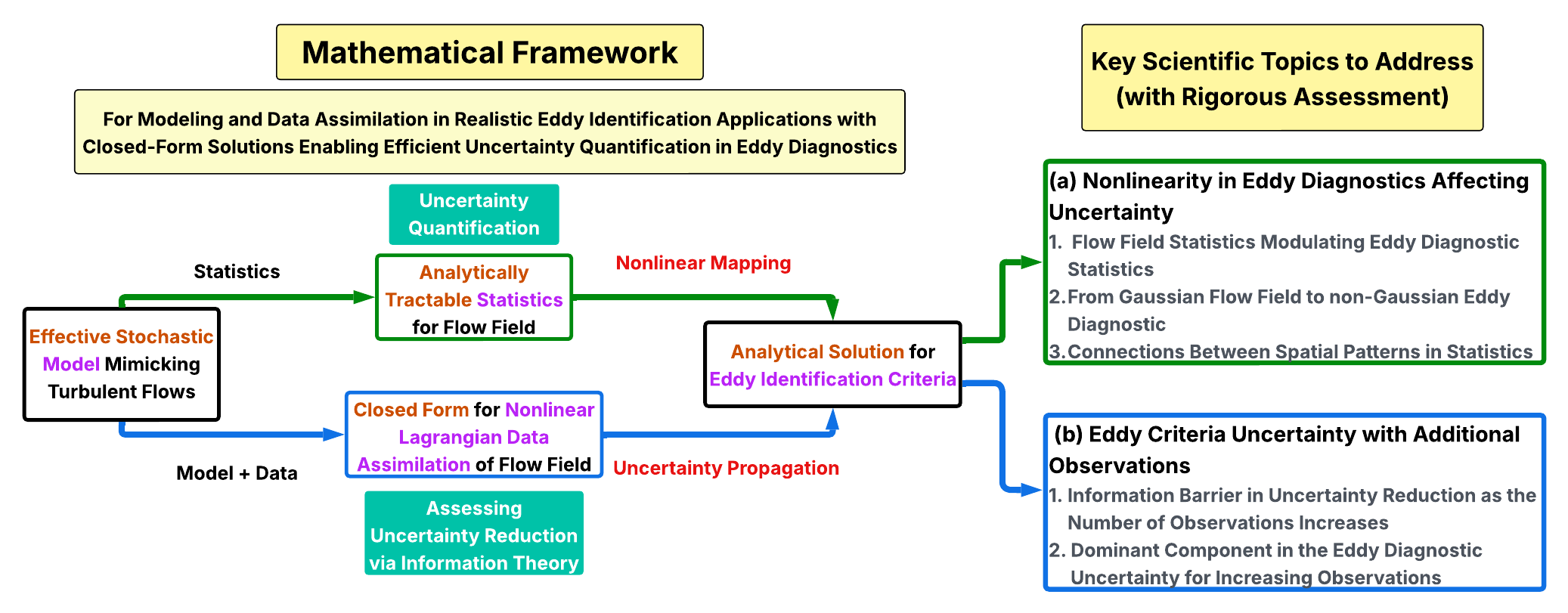}
    \caption{Schematic of the mathematical framework, outlining how the tools presented in Section \ref{sec:framework} are used together to assess the nonlinear uncertainty propagation. The framework approaches the problem from two crucial angles: i) (green) how the non-linearity in the eddy diagnostic alters the uncertainty from the flow field and ii) (blue) how the uncertainty behaves as additional observations are incorporated.}
    \label{fig:overview_schematic}
\end{figure}

The mathematical framework is summarized in Figure \ref{fig:overview_schematic}. It uses closed-form solutions to assess uncertainty propagation through the nonlinear eddy criteria from two crucial angles: i) (green arrows) how the uncertainty is affected by the nonlinearity, and ii) (blue arrows) how the uncertainty behaves as additional observations are introduced.

The building block of the framework is a computationally efficient stochastic model, which mimics turbulent flow features while maintaining analytically tractable model statistics (Section \ref{sec:FF_model}). First, the closed-form solutions for the flow field statistics are nonlinearly mapped to the eddy diagnostics, providing explicit relations between the flow field and eddy criteria statistics. These analytic formulae allow rigorous and efficient uncertainty quantification for both the flow field and the eddy diagnostic, bypassing computationally intractable ensemble methods. Second, the stochastic flow field model also serves as the forecast model in a nonlinear yet analytically tractable Lagrangian data assimilation scheme (Section \ref{sec:LaDA_scheme}). The closed-form statistics of the estimated states from the Lagrangian data assimilation allow for efficient uncertainty quantification and facilitate the study of uncertainty behavior for an increasing number of observations using information theory (Section \ref{sec:Info}).

Utilizing these tools, the framework rigorously addresses several crucial topics. From the first perspective, it examines how nonlinearity alters uncertainty. The framework explicitly details how flow field statistics modulate those of the eddy diagnostic, how the full flow field distribution is transformed by the nonlinearity, and allows analytic study of the connections between the spatial patterns of the statistics. From the second perspective, it assesses the uncertainty behavior of both the flow field and eddy diagnostics when observations are incorporated. It identifies practical information barriers in uncertainty reduction and the dominating components of the uncertainty as more observations are available.

\subsection{Stochastic model for the random flow field}\label{sec:FF_model}

The root of the framework lies in the simple stochastic model used to simulate the underlying ocean dynamics. It manages to imitate key turbulent features while maintaining analytically tractable statistics for the flow field. Consider the two-dimensional velocity field $\mathbf{u}(\mathbf{x},t) = [u,v]^\mathtt{T} $ on a double periodic domain $(0, 2\pi]^2$, with the coordinates $\mathbf{x} = [x,y]^\mathtt{T}$. Since the boundary conditions are periodic, we can approximate the velocity field using a truncated sum of its Fourier bases \cite{chen2014information, chen2024lagrangian, majda2016introduction},
\begin{equation}\label{eq:u_fourier}
\mathbf{u}(t, \mathbf{x}) = \sum_{\mathbf{k} \in \mathcal{K}} \hat{u}_{\mathbf{k}}(t) \cdot e^{i\mathbf{k \cdot x}} \cdot \mathbf{r_k},
\end{equation}
which is also known as its spectral representation.
Here, $\hat{u}_{\mathbf{k}}(t)$ is the Fourier time series for the mode $\mathbf{k} = [k_1,k_2]^\mathtt{T} \in \mathcal{K}$, $e^{i\mathbf{k \cdot x}}$ is the 2-D Fourier basis function, and $\mathbf{r_k}=(r_{\mathbf{k},1}, r_{\mathbf{k},2})^\mathtt{T}$ is the eigenvector associated with wave number $\mathbf{k}$ which connects the two velocity components. In turbulent systems, the nonlinearity enables energy transfer between different modes, resulting in a complex model for each Fourier coefficient $\hat{u}_{\mathbf{k}}(t)$. Instead of explicitly including the interactions between modes, here each Fourier coefficient $\hat{u}_{\mathbf{k}}(t)$ follows a complex Ornstein-Uhlenbeck (OU) process \cite{gardiner1985handbook},
\begin{equation}\label{eq:dvk}
\frac{\text{ d} \hat{u}_{\mathbf{k}}(t)}{\text{ d}t} = ( - d_{\mathbf{k}} + i \phi_{\mathbf{k}}) \hat{u}_{\mathbf{k}}(t) + f_{\mathbf{k}}(t) + \sigma_{\mathbf{k}}\dot{W}_{\mathbf{k}},
\end{equation}
where $i$ is the imaginary unit. The real-valued constant parameters $d_k > 0$, $\phi_{\mathbf{k}}$, and $\sigma_{\mathbf{k}}$ are the damping rate, phase speed, and noise amplitude, respectively. The complex-valued forcing $f_{\mathbf{k}}(t)$ can either be a constant or a time-dependent deterministic function, and $W_{\mathbf{k}}$ is a complex Wiener process. For presentation convenience, we set $f_{\mathbf{k}}(t) \equiv f_{\mathbf{k}}$ to a constant in the following discussion. When $f_{\mathbf{k}}(t)$ is a time-periodic function, often used to represent seasonal or annual cycles, a similar framework can be applied. The only difference is that the steady-state mean value becomes a periodic function instead of a constant. In the incompressible case, we also have that \cite{chen2023stochastic},
\begin{equation}\label{eq:rk_incompressible}
 \mathbf{r_{k\neq 0}} = \begin{bmatrix}
\frac{-ik_2}{\sqrt{k_1^2 +k_2^2}}\\
\frac{ik_1}{\sqrt{k_1^2 +k_2^2}}
\end{bmatrix},
\end{equation}
and two independent OU processes are used to characterize the background flow components along $x$ and $y$ directions, which are associated with mode $\mathbf{k}=(0,0)^\mathtt{T}$.

The damping, phase, and deterministic forcing in \eqref{eq:dvk} arise naturally from the dissipation, dispersion, and large-scale forcing terms when decomposing a PDE. The stochasticity (and additional damping if applicable) in \eqref{eq:dvk} compensates for the turbulent dynamics produced by deterministic nonlinearities, allowing the model to closely approximate key turbulent features. Specifically, the damping and stochastic terms account for energy loss and spontaneous gain between modes \cite{majda2016introduction, majda1999models, majda2001mathematical, majda2002priori}. For this reason, such a dynamical structure is commonly used as a forecast model to accelerate data assimilation for turbulent systems \cite{covington2025probabilistic, farrell1993stochastic, berner2017stochastic, harlim2008filtering, branicki2013non}. Thus, this stochastic model provides a sufficient approximation for eddy identification in this context. While the model effectively captures the key turbulent dynamics, the simple structure facilitates theoretical study of uncertainty propagation as the statistics have known closed formulae \cite{gardiner1985handbook, srkk2013bayesian},
\begin{eqnarray}
\label{eq:FF_prior_mean}\mu_\mathbf{k}^{att} &=& \frac{f_\mathbf{k}}{ d_{\mathbf{k}} - i \phi_{\mathbf{k}}},\\
\label{eq:FF_prior_var}R_{\mathbf{k}}^{att} &=& \frac{\sigma^2_\mathbf{k}}{2 d_\mathbf{k}}.
\end{eqnarray}
Since the flow field model \eqref{eq:dvk} is linear with additive noise, its statistical steady-state distribution, also known as the prior distribution, is Gaussian. It is the most natural way to statistically estimate the flow state solely using the model. Due to the Gaussianity, the prior distribution is fully determined by the first two central moments \eqref{eq:FF_prior_mean} and \eqref{eq:FF_prior_var}. Thus, the closed formulae \eqref{eq:FF_prior_mean} and \eqref{eq:FF_prior_var} enable theoretical study using the full flow field distribution while bypassing high-cost Monte Carlo simulations used in most other eddy uncertainty studies. While ensemble methods can capture the flow field statistics relatively well with moderate cost, the ensemble error is heavily amplified by nonlinearity in computing the higher order statistics of eddy diagnostics, even the variance (which will be shown below), making it an impractical tool for uncertainty quantification in eddy identification. The framework developed here allows efficient numerical analysis of the nonlinear uncertainty propagation. Specifically, the closed formulae will enable the flow field statistics to inform eddy diagnostic statistics through nonlinear mapping directly.

\subsection{Nonlinear Lagrangian data assimilation}\label{sec:LaDA_scheme}

The stochastic flow field model presented in Section \ref{sec:FF_model} can be used alone to estimate the flow field with an associated uncertainty indicated by the prior distribution. To improve the flow field estimate and reduce uncertainty, the framework incorporates observations into the model simulation using data assimilation to produce what is called the posterior distribution. Specifically, the framework allows the development of a nonlinear, but analytically tractable, Lagrangian data assimilation scheme to integrate information from increasing numbers of observations, which facilitates the study of uncertainty reduction. The Lagrangian data assimilation uses the displacements of Lagrangian tracers $\mathbf{x}_l = [x_l, y_l]^\mathtt{T}$, which follow the fluid flow, to adjust the model solution and minimize the uncertainty. Since velocity is the derivative of displacement, we can define the observational process of the $l^{\text{th}}$ tracer as \cite{chen2023stochastic},
\begin{equation}\label{eq:obs_process}
    \frac{\text{ d} \mathbf{x}_l (t)}{\text{ d}t} = \mathbf{u}(t, \mathbf{x}) + \sigma_x \dot{\mathbf{W}}_l^{\mathbf{x}}(t) = \sum_{k \in \mathcal{K}} \hat{u}_{\mathbf{k}}(t) \cdot e^{i\mathbf{k \cdot x}} \cdot \mathbf{r_k}  + \sigma_x \dot{\mathbf{W}}_l^{\mathbf{x}}(t),
\end{equation}
where $\sigma_x$ is the noise coefficient, and $ \mathbf{W}_l^{\mathbf{x}}(t)$ is a Wiener process. The stochasticity accounts for unresolved modes not included in the set $\mathcal{K}$. Note that since the displacement $\mathbf{x}$ appears in the exponential function on the right-hand side of \eqref{eq:obs_process}, the observational process is highly nonlinear. Lagrangian data assimilation is widely used in practice. Examples of Lagrangian observations include ARGO drifters \cite{jayne2017argo}, sea ice trackers \cite{lopez2019ice}, and atmospheric balloons \cite{businger1996balloons}. Many of these observations have been used for eddy identifications.

Now, to define the coupled Lagrangian data assimilation system, let $\mathbf{X}(t) = [x_1, y_1, x_2,y_2 \ldots]^\mathtt{T}$ denote the vector containing the $L$ tracers' displacements $\mathbf{x}_l(t)$ and $\mathbf{U}(t)$ the vector containing the $K$ Fourier coefficients $\hat{u}_{\mathbf{k}}(t)$ at time $t$. This allows the system in \eqref{eq:u_fourier} and \eqref{eq:obs_process} to be summarized as,
\begin{eqnarray}
    \label{eq:LDA_obs_process} \frac{\text{ d} \mathbf{X}(t)}{\text{ d}t} &= & \mathbf{P_X}(\mathbf{X}(t)) \mathbf{U}(t)  + \mathbf{\Sigma}_{\mathbf{X}} \mathbf{\dot{W}_X},\\
    \label{eq:LDA_unobs_process} \frac{\text{ d}\mathbf{U}(t)}{\text{ d}t} & = & - \mathbf{\Gamma}\mathbf{U}(t) + \mathbf{F}(t)  + \mathbf{\Sigma_U} \mathbf{\dot{W}_U},
\end{eqnarray}
where $\mathbf{P_X}(\mathbf{X})$ is a matrix containing the Fourier basis functions multiplied by $r_{\mathbf{k},1}$ for $x$ displacements and $r_{\mathbf{k},2}$ for $y$ displacements. The matrices $\mathbf{\Sigma}_{\mathbf{X}}$, $\mathbf{\Gamma}$, and $\mathbf{\Sigma_U}$ are diagonal with elements $\sigma_{\mathbf{k}}$, $d_{\mathbf{k}} + i\phi_{\mathbf{k}}$, and $\sigma_x$, respectively. Since the observational process is highly nonlinear, the model is able to mimic realistic turbulent dynamics. Nevertheless, since the flow field model is linear, when observations (one realization of $\mathbf{X}$) are known, the whole system becomes linear. Thus, equations \eqref{eq:LDA_obs_process} and \eqref{eq:LDA_unobs_process} form a conditional Gaussian nonlinear system (CGNS). For such a system, the posterior distribution, which is the conditional distribution of the flow state given one realization of the observed tracer trajectories, is Gaussian and has closed-form known solutions. Particularly, the conditional distribution, known as the filter distribution, $p(\mathbf{U}(t,\mathbf{k})| \textbf{X}(s \leq t))$, has mean and covariance \cite{liptser2013statistics, chen2018conditional},
\begin{eqnarray}
    \label{eq:mu_f_LDA} \text{ d}\boldsymbol{\mu}_f &=& (\mathbf{F}(t) -\mathbf{\Gamma} \boldsymbol{\mu}_f)\text{d}t + \left(\mathbf{R}_f \mathbf{P_X}(\mathbf{X}(t))^*\right)(\mathbf{\Sigma}_{\mathbf{X}} \mathbf{\Sigma}_{\mathbf{X}}^*)^{-1}(\text{d}\mathbf{X}(t) - \mathbf{P_X}(\mathbf{X}(t)) \boldsymbol{\mu}_f \text{d}t),\\
    \label{eq:R_f_LDA} \text{ d}\mathbf{R}_f &=& [-\mathbf{\Gamma} \mathbf{R}_f - \mathbf{R}_f\mathbf{\Gamma}^* + \mathbf{\Sigma_U}\mathbf{\Sigma_U}^* - (\mathbf{R}_f\mathbf{P_X}(\mathbf{X}(t))^*) (\mathbf{\Sigma}_{\mathbf{X}} \mathbf{\Sigma}_{\mathbf{X}}^*)^{-1} (\mathbf{P_X}(\mathbf{X}(t))\mathbf{R}_f)]\text{d}t,
\end{eqnarray}
where $\boldsymbol{\mu}_f$ is a vector containing the conditional filter mean of each Fourier mode $\hat{u}_{\mathbf{k}}$, and $\mathbf{R}_f$ is their filter covariance matrix. The notation $\cdot^*$ denotes the complex conjugate. The filter solution systematically corrects the model forecast using observations based on the uncertainty in the forecast relative to that of the observations. The first term in \eqref{eq:mu_f_LDA}, $(\mathbf{F}(t) -\mathbf{\Gamma} \boldsymbol{\mu}_f)\text{d}t$, is the forecast from the model, while the second term incorporates the observations. Specifically, the second term is composed of the Kalman gain $(\mathbf{R}_f\mathbf{P_X}(\mathbf{X}(t))^*) (\mathbf{\Sigma}_{\mathbf{X}} \mathbf{\Sigma}_{\mathbf{X}}^*)^{-1}$ and the correction $(\text{d}\mathbf{X}(t) - \mathbf{P_X}(\mathbf{X}(t)) \boldsymbol{\mu}_f \text{d}t)$. The Kalman gain is made up of the forecast uncertainty $\mathbf{R}_f$, the observational operator $\mathbf{P_X}(\mathbf{X}(t))^*)$, and the covariance of the observational noise $\mathbf{\Sigma}_{\mathbf{X}} \mathbf{\Sigma}_{\mathbf{X}}^*$, while the correction is the difference between the observed tracer locations $\text{d}\mathbf{X}(t)$ and their expected position based off of the model $\mathbf{P_X}(\mathbf{X}(t)) \boldsymbol{\mu}_f \text{d}t$. Note that when the forecast uncertainty is large compared to the observational uncertainty, the Kalman gain will be large, putting more weight on the observational correction. On the other hand, when the forecast uncertainty is small compared to the observation uncertainty, the Kalman gain will be small, and the update will rely more on the forecast model.

It is worth highlighting that even though different Fourier coefficients $\hat{u}_\mathbf{k}$ are independent of each other since they are driven by independent OU processes in \eqref{eq:dvk}, correlations exists in the posterior estimate via $\mathbf{P_X}$ in the time evolution of the covariance matrix \eqref{eq:R_f_LDA}. This is because the observed tracer is driven by the full flow field, which is the summation of all Fourier modes. With a limited number of tracers and random noise in the observational processes, data assimilation cannot distinguish the contribution from each Fourier mode that drives the motion of the tracers.

The filter posterior distribution incorporates observations up to the current time point $t$. As a post-processing step, smoothing is done to further improve the estimate by also including the future information in the known time interval $[0,T]$. The smoother distribution $p(\mathbf{U}(t,\mathbf{k})| \textbf{X}(s \in [0,T]))$ is also Gaussian and the statistics have the closed formulae \cite{chen2023stochastic},
\begin{eqnarray}
    \label{eq:mu_s_LDA} \overleftarrow{\text{ d}\boldsymbol{\mu}_s} &=& [-\mathbf{F}(t) + \mathbf{\Gamma\mu}_s+ (\mathbf{\Sigma}_U\mathbf{\Sigma}_U^*)\mathbf{R}_f^{-1}(\boldsymbol{\mu}_f - \boldsymbol{\mu}_s)]\text{ d}t,\\
    \label{eq:R_s_LDA} \overleftarrow{\text{ d}\mathbf{R}_s} &=& [-(-\mathbf{\Gamma} + (\mathbf{\Sigma}_U \mathbf{\Sigma}_U^*) \mathbf{R}_f^{-1}) \mathbf{R}_{s} - \mathbf{R}_{s}(-\mathbf{\Gamma}^* + (\mathbf{\Sigma}_U\mathbf{\Sigma}_U^*) \mathbf{R}_f^{-1}) +\mathbf{\Sigma}_U\mathbf{\Sigma}_U^*]\text{ d}t.
\end{eqnarray}
Here, $\boldsymbol{\mu}_s$ is the conditional smoother mean, and $\mathbf{R}_s$ is the conditional smoother covariance. The $\overleftarrow{\text{d}}$ indicates the negative of the forward difference. The smoother unbiasedly removes uncertainty from the estimate using future observations. By utilizing observations on both sides of the current time step, smoothing dynamically interpolates between observations. Since the smoother estimate integrates more information into the model simulation, it is generally more accurate. For this reason, the smoother estimate is used as the posterior estimate in the analysis here.

The resulting data assimilation approach efficiently estimates the states of a turbulent system. By providing closed-form posterior statistics, the scheme bypasses high-cost Monte Carlo simulations. It facilitates efficient uncertainty quantification of the posterior eddy diagnostic through the use of explicit nonlinear mapping. Notably, the closed analytic formulae of such a Lagrangian data assimilation framework advance the study of uncertainty reduction in both the flow field and the eddy diagnostic as the number of observations increases. Finally, while not used here, the Lagrangian data assimilation scheme also naturally fits with the Lagrangian descriptor eddy diagnostic \cite{mancho2013lagrangian}.

\subsection{Information theory for uncertainty reduction}\label{sec:Info}
Studying the uncertainty reduction in both the flow field and eddy criteria attributed to the incorporation of observations provides critical insights that can guide practical decisions in data collection. These insights help to optimize resources by quantifying the information gain from each additional observation. Furthermore, examining the uncertainty reduction can reveal other distinguishing features between the flow field and the eddy criteria produced by nonlinearity.

The flow field model alone produces the so-called prior distribution, which is the distribution at the statistical steady state \eqref{eq:FF_prior_mean}--\eqref{eq:FF_prior_var}. The prior distribution indicates the best statistical estimate of the flow state exploiting only the model. It is distinct from the posterior distribution produced by incorporating observations. Likewise, applying the eddy diagnostic to these two estimates will also produce distinct distributions. Assessment of each estimate's uncertainty reduction attributed to the observations requires a comparison of the prior and posterior distributions, the natural metric for which is relative entropy, a widely used information measurement. Alternative standard norms that measure path-wise improvement are known to underestimate the contribution of tail probabilities, while relative entropy is more accurate in the statistical sense, making it the optimal choice for this application \cite{majda2018model, kleeman2011information, delsole2004predictability}.

Relative Entropy, also known as Kullback-Leibler (KL) divergence, quantifies the uncertainty reduction, or information gain, from one distribution to another. Let $\mathbf{Z}(t)$ be an $n$-dimensional vector of the estimated state at time $t$ where each entry corresponds to a distinct spatial coordinate. In this case, $\mathbf{Z}$ represents either the flow field velocity or the eddy diagnostic. In addition, let $\mathbf{X}(t)$ be the $m$-dimensional vector of observations and $\mathbf{X}(s \subset S) = \{\mathbf{X}(s) : s \subset S\}$ the set of observations in the time interval $S$. Denote by $p(\mathbf{Z}(t,\mathbf{x}))$ the prior distribution and $p(\mathbf{Z}(t,\mathbf{x})| \mathbf{X}(s \subset S))$ the posterior distribution. Then, the uncertainty reduction from the prior to the posterior distribution is given by \cite{kullback1951information, kullback1997information},
\begin{equation}\label{eq:relative_entropy}
    \mathcal{P}(p(\mathbf{Z}(t,\mathbf{x})| \mathbf{X}(s \subset S)), p(\mathbf{Z}(t,\mathbf{x}))) = \int p(\mathbf{Z}(t,\mathbf{x})| \mathbf{X}(s \subset S)) \ln\left( \frac{p(\mathbf{Z}(t,\mathbf{x})| \mathbf{X}(s \subset S))}{p(\mathbf{Z}(t,\mathbf{x}))} \right) \d \mathbf{x}
\end{equation}
Relative entropy has two key properties. First, it is non-negative; it is only equal to zero when the two distributions are identical and grows as the distributions deviate from each other. Second, relative entropy is invariant under nonlinear transformations, meaning it is dimensionless and thus will not change across different coordinate systems or units. These make relative entropy a natural coordinate-independent measure for comparing the flow field and eddy diagnostic uncertainty reduction from incorporating observations.

The stochastic flow field model \eqref{eq:dvk} and the Lagrangian data assimilation scheme \eqref{eq:LDA_obs_process}--\eqref{eq:R_s_LDA} utilized by the framework ensure Gaussian flow field statistics. While the nonlinearity in the eddy criteria does not guarantee a Gaussian diagnostic, the framework utilizes a normal approximation as it is the least biased estimate. Then, the relative entropy of both the flow field and eddy criteria can be fully determined by the first two central moments. Specifically, for the general estimated state $\mathbf{Z}$ with prior distribution $p(\mathbf{Z}(t,\mathbf{x})) \sim \mathcal{N}(\mathbf{m}^{att}(t), \mathbf{R}^{att}(t))$ and posterior distribution $p(\mathbf{Z}(t,\mathbf{x})| \mathbf{X}(s \subset S)) \sim \mathcal{N}(\mathbf{m}(t), \mathbf{R}(t))$, the uncertainty reduction \eqref{eq:relative_entropy} becomes,
\begin{equation}\label{eq:reletive_entropy_gaus}\begin{split}
    \mathcal{P}(p(\mathbf{Z}(t,\mathbf{x})| \mathbf{X}(s \subset S)), p(\mathbf{Z}(t,\mathbf{x}))) &= \frac{1}{2}[(\mathbf{m}-\mathbf{m}^{att})^T(\mathbf{R}^{att})^{-1}(\mathbf{m}-\mathbf{m}^{att})]  \quad \quad \quad \quad \quad \quad \hspace{0.08cm} \ldots \text{ Signal}\\
    & + \frac{1}{2}[\mbox{tr}(\mathbf{R}((\mathbf{R}^{att})^{-1}) - n - \ln{\det(\mathbf{R}(\mathbf{R}^{att})^{-1})}] \quad \ldots \text{ Dispersion}.
    \end{split}
\end{equation}
Note that the explicit dependence of $\mathbf{m},\mathbf{R}, \mathbf{m}^{att}, \text{ and } \mathbf{R}^{att}$ on $t$ is left out of Equation \eqref{eq:reletive_entropy_gaus} for notation simplicity. Equation \eqref{eq:reletive_entropy_gaus} can be broken into two components to reveal specific information about the uncertainty in the mean and variance. The first term, called the `signal', measures the uncertainty reduction in the mean weighted by the prior variance, while the second term, i.e., the `dispersion', quantifies the uncertainty reduction in the variance. Relative entropy has been widely used as a tool for uncertainty quantification in the oceanic and atmospheric sciences \cite{kleeman2002measuring, delsole2004predictability, majda2010quantifying, gershgorin2012quantifying, pierard2024quantifying}.

\section{Key Definitions for Framework Demonstration}\label{sec:defs}
In the following subsections we outline the details of an eddy identification problem to demonstrate framework's application.

\subsection{Eddy identification criteria}

The objective of this study is to demonstrate how the mathematical framework developed here can be utilized to improve the understanding of eddy identification in the presence of uncertainty. To that end, the framework is tested on the most widely used eddy identification criteria: the Okubo-Weiss (OW) parameter \cite{chelton2007global, cheng2014statistical, isern2003identification, isern2006vortices, petersen2013three}. Nevertheless, it is important to note that the framework is general and can be applied to various eddy diagnostics beyond the OW parameter.

The OW parameter is used to detect eddies at a fixed instant in time using snapshot data. It is given by \cite{okubo1970horizontal,weiss1991dynamics},
\begin{equation}\label{eq:OW}
    \text{OW}(\mathbf{u}) = s_n^2 + s_s^2 - \omega^2,
\end{equation}
\begin{equation}\label{eq:strain_vort}
    s_n = u_x - v_y, \quad s_s = v_x +u_y, \quad \text{and} \quad \omega = v_x - u_y,
\end{equation}
where $s_n$ and $s_s$ are the normal and shear strains, while $\omega$ is the relative vorticity. The subscripts of $u$ and $v$ indicate the derivative. Because the flow field in computing the OW parameter is at a fixed time instant, the time variable is omitted here. In whole, the OW parameter measures the strength of vorticity relative to the strain. Negative values of the OW parameter indicate vortical flow, and eddies are identified in regions below a given threshold value. The threshold is chosen based on the application, but the most common threshold is $-0.2 \sigma_{OW}$, where $\sigma_{OW}$ is the spatial standard deviation of the OW parameter.

In the case of incompressible flow, $u_x = -v_y$, and thus the OW parameter can be further simplified to,
\begin{equation}\label{eq:OW_incompressible}
      \text{OW}(\mathbf{u}) = 4u_x^2 + 4v_xu_y.
\end{equation}
For the ocean surface velocity field, incompressible flow is a sufficient approximation. Thus, the definition given in \eqref{eq:OW_incompressible} will be utilized in the following application.

\subsection{Dynamical regime}\label{sec:dyn_regime}

The primary goal of the mathematical framework is to provide a theoretical analysis of the uncertainty in eddy identification. However, numerical simulations are also run to give complementary insights. For all test cases (except the validation tests to reduce the computational cost using the Monte Carlo simulation), we employ modes spanning $[-4,4]^2$ with equipartitioned energy, which is a configuration that reflects essential turbulent features \cite{alexakis2018cascades, bell2022thermal, cichowlas2005effective}. Only the non-zero modes are used in this study, as the anomalous dynamics are more important in the eddy identification problem. Particularly in the case of the OW parameter, the diagnostic depends entirely on the $x$ and $y$ derivatives of $u$ and $v$, for which the background modes make no contribution, as they are constant across space.

The parameter values used in numerical simulations are $\sigma_x = 0.1$, $d_{\mathbf{k}} = 0.5$, $\sigma_{\mathbf{k}} = 0.15$, $\phi_{\mathbf{k}}=0$, and $f_{\mathbf{k}}(t) = 0$, unless otherwise noted. Additionally, incompressibility is enforced, and thus \eqref{eq:rk_incompressible} and \eqref{eq:OW_incompressible} hold. For numerical integration, we use a time step size of $\Delta t = 0.001$ and run the simulation for a total time of $T = 10$. The initial conditions for \eqref{eq:LDA_obs_process} are randomly chosen from a uniform distribution on $(0,2\pi]^2$, while those for \eqref{eq:LDA_unobs_process} are set to 0. The initial condition for the filter mean is the true signal at $t=0$, which is 0, and the filter covariance initial condition is a diagonal matrix with entries $0.0001$. The smoother initial conditions are the filter mean and covariance at time $t=T=10$.

\section{Uncertainty Quantification in Eddy Identification Criteria}\label{sec:Eddy_UQ}

\subsection{Mean and variance of the OW parameter}
Since the flow field is turbulent and is modeled by stochastic processes, the flow velocities $u$ and $v$ are random variables at each fixed time instant. Let us apply the mean-fluctuation decomposition to the flow variables \cite{adrian2000analysis},
\begin{eqnarray}\label{eq:mean_fluc_decomp}
    u = \bar{u} + u'\qquad\mbox{and}\qquad v = \bar{v} + v',
\end{eqnarray}
where $\bar{\cdot}$ denotes the statistical average at a fixed time instant over the random variable, which is also known as the ensemble average. Therefore, $\bar{u} = \mbox{E}(u)$ and $\bar{v} = \mbox{E}(v)$ are constants while $u'$ and $v'$ remain random variables but with mean $0$.

The mean and variance of the OW parameter can be computed by plugging the mean-fluctuation decomposition of the flow variables \eqref{eq:mean_fluc_decomp} into the OW parameter \eqref{eq:OW_incompressible}, and then applying the expectation or variance. The following propositions present the key analytic formulae derived. The details are provided in Section \ref{Asec:deriv_E(OW)_Var(OW)} of the Appendix.

\begin{prop}[Expectation of the OW parameter]\label{prop:E(OW)}

Given the two-dimensional incompressible random velocity field $\mathbf{u}(t,\mathbf{x}) = [u,v]^\mathtt{T}$, satisfying the spectral decomposition \eqref{eq:u_fourier} with different Fourier coefficients being independent of each other, and the mean-fluctuation decomposition \eqref{eq:mean_fluc_decomp}, the expectation of the OW parameter \eqref{eq:OW_incompressible} is given by,
\begin{equation}\label{eq:E(OW)_incompres_PWA}
    \mbox{E}(\mbox{OW}(\mathbf{u})) = 4 (\bar{u}_x^2 + \bar{v}_x\bar{u}_y).
\end{equation}
\end{prop}

The expectation of the OW parameter \eqref{eq:E(OW)_incompres_PWA} is equivalent to the OW parameter applied to the expectation of the flow velocity, despite the nonlinearity in the eddy criteria. This is a direct result of the Fourier decomposition \eqref{eq:u_fourier} and the incompressibility condition. In general, the expectation of the OW parameter also includes cross terms involving the fluctuation components which cancel under the incompressible Fourier representation. Note that the result in \eqref{eq:E(OW)_incompres_PWA} applies to a general random flow field since we do not assume a specific distribution, such as a Gaussian, for the flow field statistics. See the Appendix Section \ref{Asec:deriv_E(OW)_Var(OW)} for the details. 

\begin{prop}[Variance of the OW parameter]\label{prop:Var(OW)_general}
    Consider the same two-dimensional incompressible random velocity field $\mathbf{u}(t,\mathbf{x}) = [u,v]^\mathtt{T}$ from Proposition \ref{prop:E(OW)}. Additionally, assume the flow field  is Gaussian. Given the mean fluctuation decomposition \eqref{eq:mean_fluc_decomp}, the variance of the OW parameter \eqref{eq:OW_incompressible} is,
    \begin{eqnarray} \label{eq:var(OW)_general_phy}
    \text{Var}[\text{OW}(\mathbf{u})] &=& 16\bigg[3\overline{u_x'^2}^2 + 4 \overline{u_x'v_x'} \hspace{0.2cm}\overline{u'_x u'_y}   + \overline{v_x'^2}\hspace{0.2cm}\overline{u_y'^2} \\
    \nonumber &+& 4 \overline{u_x}^2 \overline{u_x'^2} - 2 \bar{v}_x \bar{u}_y \overline{u_x'^2} + \overline{v_x}^2 \overline{u_y'^2}  + \overline{u_y}^2 \overline{v_x'^2} +  4\bar{u}_x \bar{v}_x \overline{u_x'u_y'} + 4\bar{u}_x\bar{u}_y \overline{u_x'v_x'}  \bigg].
\end{eqnarray}
 To reveal the underlying mathematical structure, \eqref{eq:var(OW)_general_phy} can equivalently be expressed using the spectral representation,
\begin{eqnarray}\label{eq:Var(OW)_Fourier_general}
    \text{Var}[\text{OW}(\mathbf{u})] &=& \sum_\mathbf{k} \sum_{\mathbf{k}'} \overline{|\hat{u}_\mathbf{k}'|^2} \hspace{0.2cm}\overline{|\hat{u}_{\mathbf{k}'}'|^2} (3a_1a_1' + 4b_1b_1' + c_1c_1')\\
    \nonumber &+& \sum_{\mathbf{k}} \sum_{\mathbf{k}'} \sum_{\mathbf{k}''} \overline{\hat{u}_\mathbf{k}} \hspace{0.2cm}\overline{\hat{u}_{-\mathbf{k}'}}\hspace{0.2cm}\overline{|\hat{u}_{\mathbf{k}''}'|^2} (4d_1d_1'd_2'' - 2f_1f_1'f_2'' + g_1g_1'g_2''\\
    \nonumber & & \hspace{4.4cm}  + h_1h_1'h_2''+ 4j_1j_1'j_2'' + 4l_1l_1'l_2'') e^{i (\mathbf{k-k'}) \cdot\mathbf{x}}.
\end{eqnarray}
Here, the coefficients from the first term are $a_i$, the second term $b_i$ etc. Each coefficient is a function of $\mathbf{k}$, $\mathbf{k}'$ or $\mathbf{k}''$. For the latter two, the coefficients are denoted with a ` $'$ ' or ` $''$  ', respectively. The functional forms of the coefficients are given in Table \ref{tab:Fourier_Var_coeffs} in the Appendix.
\end{prop}

The variance formulae in Proposition \ref{prop:Var(OW)_general} can be further refined by assuming an isotropic flow field, which means the statistics of the flow field are equivalent across direction. Isotropic turbulence is often used as a test case to study the fundamental properties of a turbulent system, as small scales realistic turbulence is isotropic \cite{batchelor1953theory, tennekes1972first, thorpe2005turbulent, farhat2024identifying}. The equal energy partition, used in the numerical simulations here, falls under this classification, thus we use the variance formulae defined in the following Proposition \ref{prop:Var(OW)_iso} for analysis in the subsequent sections.

\begin{prop}[Variance of the OW parameter for an isotropic flow field]\label{prop:Var(OW)_iso}
    Consider the same two-dimensional incompressible random velocity field $\mathbf{u}(t,\mathbf{x}) = [u,v]^\mathtt{T}$ from Proposition \ref{prop:Var(OW)_general}. Additionally, assume the flow field  has an isotropic energy distribution. Given the mean fluctuation decomposition \eqref{eq:mean_fluc_decomp}, the variance of the OW parameter \eqref{eq:OW_incompressible} is,
    \begin{eqnarray}\label{eq:var_phy_iso}
        Var[\mbox{OW}(\mathbf{u})] &=& 16\bigg[3\overline{u_x'^2}^2   + \overline{v_x'^2}\hspace{0.2cm}\overline{u_y'^2} \\
        \nonumber &+& 4 \overline{u_x}^2 \overline{u_x'^2} - 2 \bar{v_x} \bar{u_y} \overline{u_x'^2} + \overline{v_x}^2 \overline{u_y'^2}  + \overline{u_y}^2 \overline{v_x'^2} \bigg].
    \end{eqnarray}
   To reveal the underlying mathematical structure, \eqref{eq:var_phy_iso} can equivalently be expressed using the spectral representation,
    \begin{eqnarray}\label{eq:Var(OW)_Fourier_iso}
    \text{Var}[\text{OW}(\mathbf{u})] &=& \sum_\mathbf{k} \sum_{\mathbf{k}'} \overline{|\hat{u}_\mathbf{k}'|^2} \hspace{0.2cm}\overline{|\hat{u}_{\mathbf{k}'}'|^2} (3a_1a_1' + b_1b_1')\\
    \nonumber &+& \sum_{\mathbf{k}} \sum_{\mathbf{k}'} \sum_{\mathbf{k}''} \overline{\hat{u}_\mathbf{k}} \hspace{0.2cm}\overline{\hat{u}_{-\mathbf{k}'}}\hspace{0.2cm}\overline{|\hat{u}_{\mathbf{k}''}'|^2} (4d_1d_1'd_2'' - 2f_1f_1'f_2'' + g_1g_1'g_2'' + h_1h_1'h_2'') e^{i (\mathbf{k-k'}) \cdot\mathbf{x}}
\end{eqnarray}
    Note that the coefficients in \eqref{eq:Var(OW)_Fourier_iso}, are the same as those in Equation \eqref{eq:Var(OW)_Fourier_general}, which are recorded in Table \ref{tab:Fourier_Var_coeffs} of the Appendix. The terms corresponding to the coefficients $c$, $j$, and $l$ cancel as a result of the isotropic condition.

\end{prop}

  The closed expectation and variance formulae \eqref{eq:E(OW)_incompres_PWA}--\eqref{eq:Var(OW)_Fourier_iso} reveal explicit relationships between the flow field and eddy diagnostic statistics. It follows directly from the physical space representation that the expectation of the OW parameter \eqref{eq:E(OW)_incompres_PWA} will be fully determined by the mean spectral state of the flow field, as the spatial derivatives in Fourier space translate to multiplication by a constant. However, this is not immediately clear in the physical form of the variance for either the general energy distribution \eqref{eq:var(OW)_general_phy}, or the isotropic \eqref{eq:var_phy_iso}. The Fourier representations of the variance \eqref{eq:Var(OW)_Fourier_general} and \eqref{eq:Var(OW)_Fourier_iso} show that it is fully determined by the first two central moments of the flow field, $\overline{\hat{u}_\mathbf{k}}$ and $\overline{|\hat{u}_\mathbf{k}'|^2}$. As a result, the variance of the OW parameter can be computed directly from the prior or posterior flow field estimates. Also note that both equations \eqref{eq:Var(OW)_Fourier_general} and \eqref{eq:Var(OW)_Fourier_iso} are quadratic functions of the flow field variance.

  In addition, the Fourier representations highlight two underlying structures of the terms in physical space: terms that contain only fluctuation components, and terms that contain two fluctuation components and two mean components, referred to here as fluctuation-fluctuation and mean-fluctuation terms, respectively. The former category depends solely on the variance of the flow field in Fourier space $\overline{|\hat{u}_\mathbf{k}'|^2}$ while the latter is also influenced by the mean $\overline{\hat{u}_\mathbf{k}}$, which incorporates dynamics into the OW uncertainty not present in the flow field. Particularly, due to the use of the global Fourier basis functions, the uncertainty in the flow field is spatially homogeneous, while the mean varies in space. Thus, the variance of the OW parameter will exhibit spatial inhomogeneity not seen in the flow field. In fact, the mean-fluctuation terms are closely linked to the expectation of the OW parameter \eqref{eq:E(OW)_incompres_PWA}. Specifically, the first two mean-fluctuation terms of \eqref{eq:var_phy_iso} and \eqref{eq:var(OW)_general_phy} can be rewritten as $2 \overline{u_x'^2}(2\overline{u_x}^2 - \bar{v}_x \bar{u}_y)$ which closely resembles the expectation formula. The remaining two mean-fluctuation terms in the isotropic variance \eqref{eq:var_phy_iso} can also be linked to the expectation of the OW parameter. These connections will be thoroughly explored in Section \ref{sec:spatial_patterns}.

\begin{figure}[h!]
    \centering
    \includegraphics[width=\linewidth]{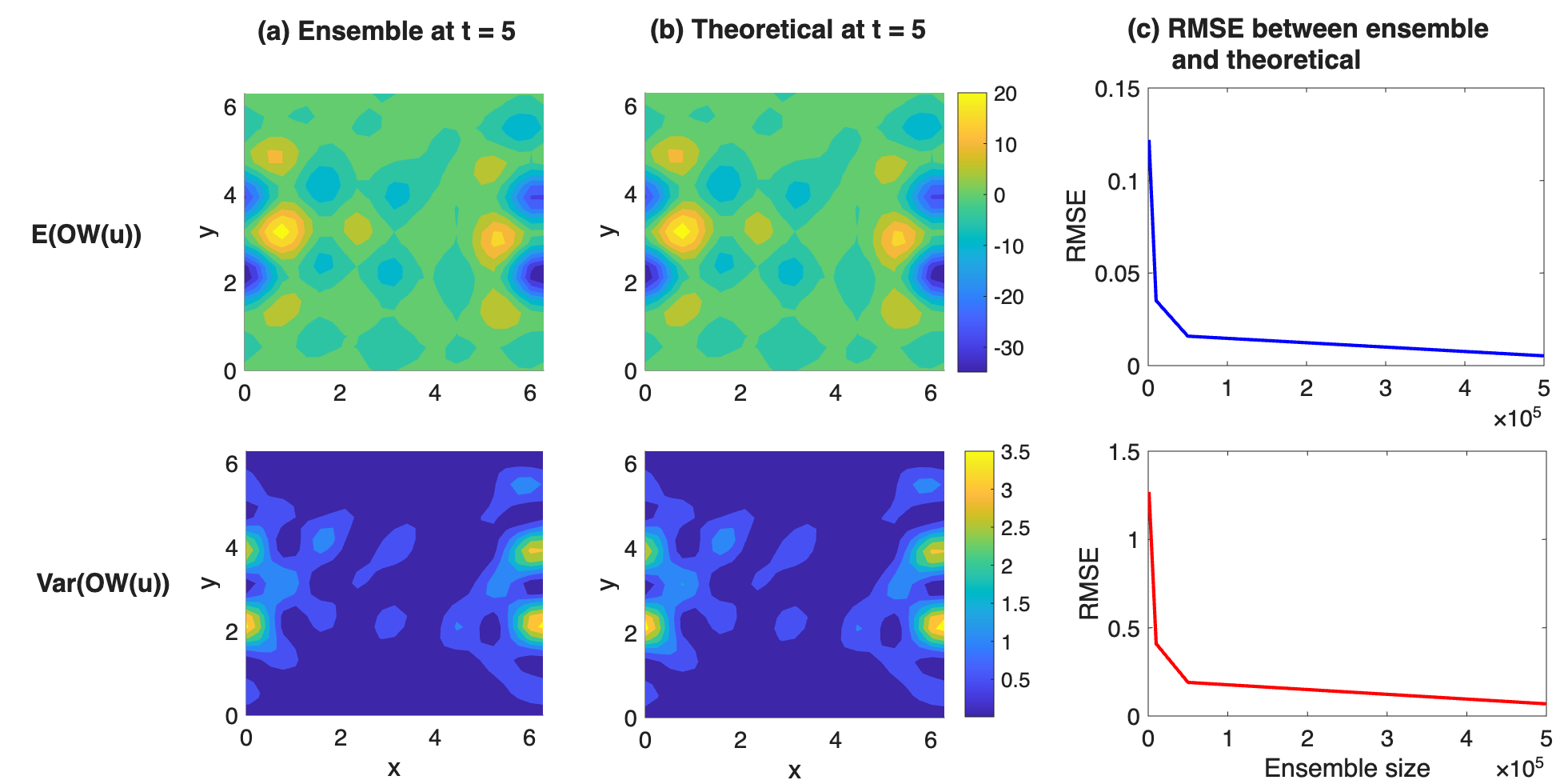}
    \caption{Comparison of the OW parameter expectation and variance computed from an ensemble and using the analytic formulae given in equations \eqref{eq:E(OW)_incompres_PWA} and \eqref{eq:var_phy_iso}. Panels (a) and (b) show the spatial snapshots of the posterior expectation and variance at $t=5$. Panel (a) is computed using an ensemble of size 500,000, while Panel (b) uses the theoretical formulae. Panel (c) shows the RMSE between the ensemble and theoretical statistics for the prior OW as the ensemble size increases. Here, the flow field contains modes $k_1,k_2 \in [-2,2]$ with the exception of the background modes.}
    \label{fig:OW_stat_validation}
\end{figure}

Figure \ref{fig:OW_stat_validation} compares the OW statistics computed numerically using an ensemble to their theoretical values using the formulae given in equations \eqref{eq:E(OW)_incompres_PWA} and \eqref{eq:var_phy_iso}, showing that, while slowly, the numerical approximations do converge to the theoretical values. In Panel (a), a large ensemble size of 500,000 is used to compute the expectation and variance of the posterior OW parameter at time $t=5$. The spatial patterns of the numerical statistics (Panel (a)) match those of the theoretical statistics (Panel (b)). Note that the y-axis of the RMSE between the numerical and theoretical expectation has one-tenth of the range of the variance (Panel (c)). The ensemble error is significantly exaggerated in the variance of the OW parameter, making ensemble methods for assessing uncertainty extremely expensive, as they require huge ensemble sizes to accurately represent the uncertainty.

\subsection{Full distribution of the OW parameter}

%\textcolor{gray}{Briefly compare the full distribution and the Gaussian approximation. Explain why Gaussian approximation is sufficient in the study here.}

%{\blue Figure X:} \textcolor{gray}{Compare the full prior distribution and the Gaussian approximation. We may also add the posterior distribution and its temporal evaluation and/or spatial distributions of the statistics (like report figures 5 and 6 but for posterior rather than the prior) }

%\textcolor{blue}{I am working on code for the posterior distribution. Since it varies in time and space, this takes more time to run. I will run it for a shorter time series, which will still give us an idea of temporal patterns, but will give the results a bit quicker.}

\begin{figure} [h!]
    \centering
    \includegraphics[width=0.5\linewidth]{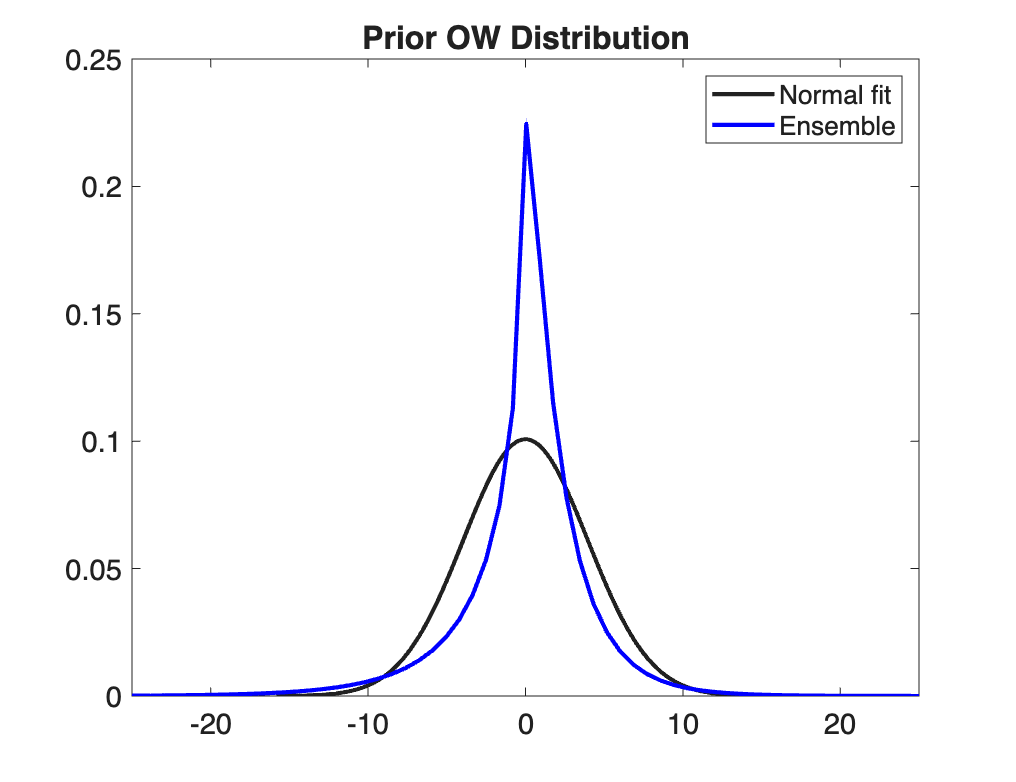}
    \caption{Comparison of the true PDF of the OW parameter computed from an ensemble (blue) with the normal approximation of equal mean and variance (black). Here, 500,000 ensemble members are used for the flow field modes $k_1,k_2 \in [-2,2]$ with the exception of the background modes.}
    \label{fig:Prior_OW_dist}
\end{figure}

 Since the stochastic model \eqref{eq:dvk} is linear with additive noise, the velocities in physical space and their spatial derivatives are jointly Gaussian. However, this does not guarantee a Gaussian OW distribution. Figure \ref{fig:Prior_OW_dist} depicts the full prior distribution of the OW parameter computed from an ensemble in blue, and its normal approximation in black. Despite a Gaussian flow field, the full distribution of the OW parameter is highly non-Gaussian due to the nonlinearity. Specifically, $s_n$, $s_s$, and $\omega$ in Equation \eqref{eq:OW} are Gaussian, and the OW parameter is a quadratic function of Gaussian random variables, giving it a weighted $\chi ^2$ distribution. Not only does this produce a fat-tailed distribution, but it can also be asymmetric with a strong skew. Under the dynamical regime presented in Section \ref{sec:dyn_regime}, the OW distribution has a strong negative skew.

While the OW parameter is non-Gaussian, to quantify the uncertainty using the relative entropy formula \eqref{eq:reletive_entropy_gaus}, we assume that it is Gaussian. The Gaussian distribution is the maximum entropy or least biased distribution under the given moment constraints. Thus, it is an appropriate baseline to gauge the uncertainty in this case. This assumption provides insights into the uncertainty based solely on the mean and variance of the OW parameter, for which we have derived formulae.

\section{Uncertainty Reduction as a Function of Number of Observations}\label{sec:uncertain_reduce_tracers}

In this Section, the uncertainty behavior of both the flow field and the OW parameter is examined as an increasing number of observations are incorporated into the data assimilation system.

\subsection{Asymptotic analysis of the posterior distribution}

To study the uncertainty behavior of the flow field and the OW parameter as an increasing number of observations are used, we aim to conduct an asymptotic analysis of the posterior estimate. We start with the flow field, as the OW parameter uncertainty is completely determined by the flow field statistics. The filter covariance matrix for the system \eqref{eq:LDA_obs_process}--\eqref{eq:LDA_unobs_process} has previously been shown to converge to a diagonal as the number of tracers becomes large \cite{chen2023uncertainty}. The entries of which can be approximated by a function of the number of tracers. It then follows that the smoother covariance converges similarly to a diagonal, with approximate entries given by the function presented in Proposition \ref{prop:Asym_var_FF}. The details of this proof are provided in the Appendix Section \ref{Asec:deriv_asym_Var(FF)_Var(OW)}.

\begin{prop}[Asymptotic posterior variance of the flow field]\label{prop:Asym_var_FF}

Consider the system given by equations \eqref{eq:LDA_obs_process}--\eqref{eq:LDA_unobs_process} with the filter estimate \eqref{eq:mu_f_LDA}--\eqref{eq:R_f_LDA} and smoother estimate \eqref{eq:mu_s_LDA}--\eqref{eq:R_s_LDA}. For the number of observations $L\gg 1$, the smoother covariance $\mathbf{R}_s$ converges to a diagonal matrix with diagonal entries corresponding to the mode $\mathbf{k}$ being,
\begin{equation}\label{eq:assym_smoother_var}
    R_{s\mathbf{k}} = \frac{\sigma_\mathbf{k}^2 }{2\sqrt{d_\mathbf{k} +  L \sigma_x^{-2}\sigma_\mathbf{k}^2}}.
\end{equation}
Intuitively, the variance of each mode $R_{s\mathbf{k}}$ decreases as more observations are incorporated. In addition, the diagonal structure implies that the Fourier modes will eventually become independent of each other in the posterior estimate, as in the prior estimate, with sufficient observational information.
\end{prop}

\begin{figure}[h!]
    \centering
    \includegraphics[width=\linewidth]{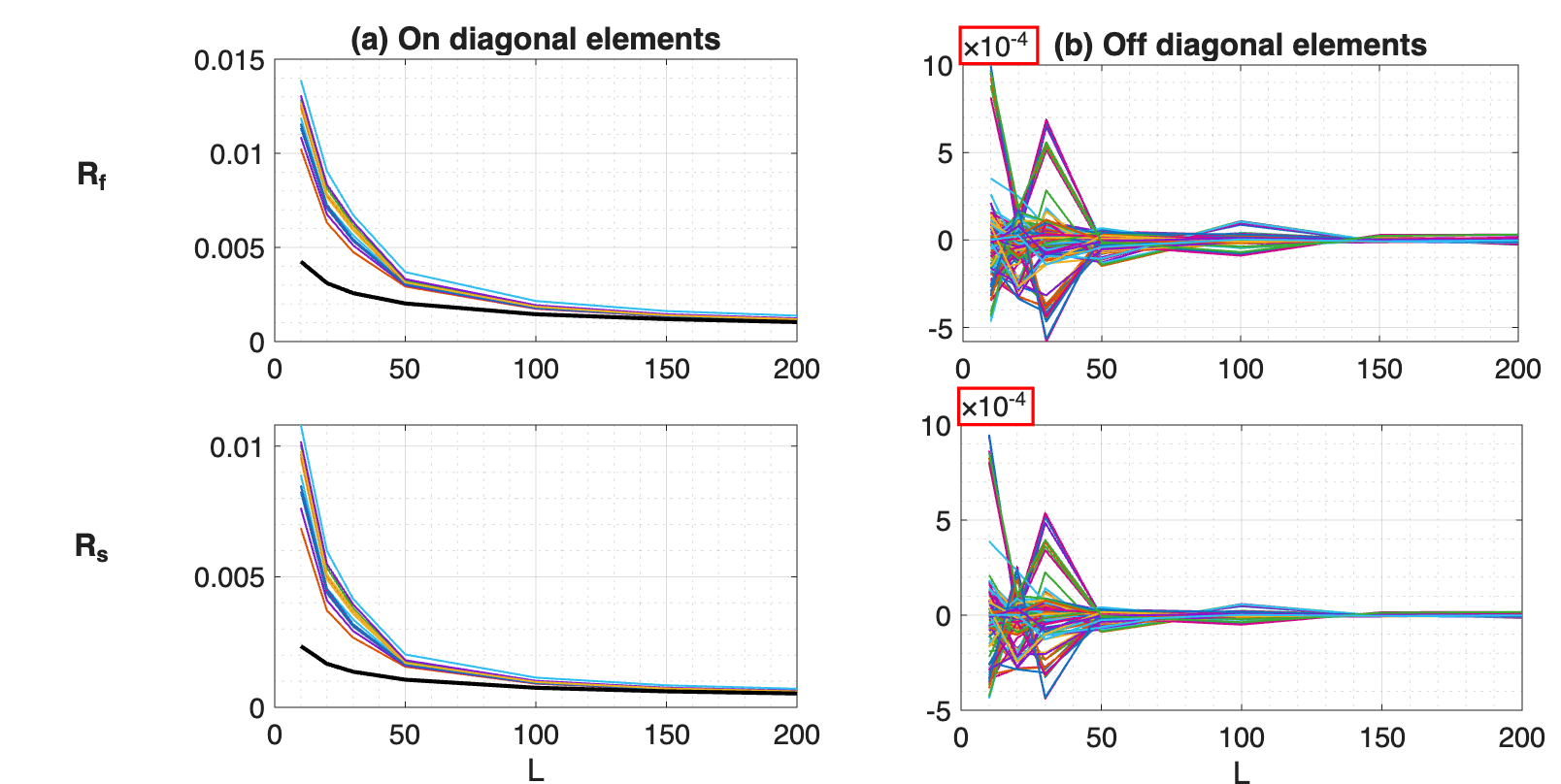}
    \caption{Time averaged values of the on (Panel (a)) and off (Panel (b)) diagonal elements of the filter (top row) and smoother (bottom row) matrices as a function of the number of observations. The colored lines are the exact values of each mode from Lagrangian data assimilation. The black line in Panel (a) is the theoretical asymptotic value of the on diagonal elements. Due to the equal energy partition outlined Section \ref{sec:dyn_regime}, this is the same for all of the modes.}
    \label{fig:Asym_post_FF_Var}
\end{figure}

Figure \ref{fig:Asym_post_FF_Var} shows the exact values of the on (Panel (a)) and off (Panel (b)) diagonal elements of the posterior variance matrices for the dynamical regime outlined in Section \ref{sec:dyn_regime} as a function of the number of observations. The off-diagonal elements of both the filter (top row) and smoother (bottom row) covariances converge to their asymptotic approximations (black) as $L$ becomes large. Furthermore, the off-diagonal entries shown in panel (b) converge to zero. Note that the y-axis in panel (b) is on the order of $10^{-4}$, and so even for a few tracers, the covariances between modes are small. Thus, the diagonal matrix approximations are reasonable estimates for the asymptotic flow field posterior variance.

The asymptotic approximation of the flow field smoother variance \eqref{eq:assym_smoother_var} can be used to obtain that of the OW parameter variance. Specifically, the results from Propositions \ref{prop:Var(OW)_general} and \ref{prop:Asym_var_FF} directly lead to Proposition \ref{prop:Asym_var_OW}.

\begin{prop}[Asymptotic posterior variance of the OW parameter]\label{prop:Asym_var_OW}

    Consider the same Lagrangian data assimilation scheme from Proposition \ref{prop:Asym_var_FF}. By direct substitution of the asymptotic estimate of the smoother covariance \eqref{eq:assym_smoother_var} into the OW variance formula \eqref{eq:Var(OW)_Fourier_general}, we obtain the asymptotic estimate for the OW parameter,
\begin{eqnarray}\label{eq:asym_var(OW)_general}
        \text{Var}[\text{OW}(\mathbf{u}_{asym})] &=& \frac{1}{4}\sum_\mathbf{k} \sum_{\mathbf{k}'} \left(\frac{\sigma_{\mathbf{k}}^2 }{\sqrt{d_{\mathbf{k}} +  L \sigma_x^{-2}\sigma_{\mathbf{k}}^2}}\right) \left( \frac{\sigma_{\mathbf{k}'}^2 }{\sqrt{d_{\mathbf{k}'} +  L \sigma_x^{-2}\sigma_{\mathbf{k}'}^2}}\right)  (3a_1a_1' + 4b_1b_1' + c_1c_1')\\
        \nonumber &+& \frac{1}{2}\sum_{\mathbf{k}} \sum_{\mathbf{k}'} \sum_{\mathbf{k}''}  \overline{\hat{u}_\mathbf{k}}\hspace{0.2cm}\overline{\hat{u}_{-\mathbf{k}'}}\left(\frac{\sigma_{\mathbf{k}''}^2 }{\sqrt{d_{\mathbf{k}''} +  L \sigma_x^{-2}\sigma_{\mathbf{k}''}^2}}\right)(4d_1d_1'd_2'' - 2f_1f_1'f_2'' + g_1g_1'g_2''\\
        \nonumber & & \hspace{7cm}  + h_1h_1'h_2''+ 4j_1j_1'j_2'' + 4l_1l_1'l_2'') e^{i (\mathbf{k-k'}) \cdot\mathbf{x}}.
    \end{eqnarray}
    The estimate for an isotropic flow field is similar, but without the terms $c$, $j$, and $l$.
\end{prop}

The asymptotic approximation \eqref{eq:asym_var(OW)_general} reveals the dominant behavior of the OW parameter uncertainty as $L$ becomes large. Since, $R_{s \mathbf{k}}$ is on the order of $1/\sqrt{L}$, the fluctuation-fluctuation terms $a$--$c$ will be O($1/L$), whereas the mean-fluctuation terms $d$--$l$ will be O($1/\sqrt{L}$). Thus, for sufficiently large $L$, the latter terms will dominate the uncertainty, and the flow field mean will have a strong influence on the OW parameter variance, which will be revealed in the spatial patterns explored further in Section \ref{sec:spatial_patterns}.

\subsection{Uncertainty reduction in the flow field and in the OW parameter}
The asymptotic variance approximations can be used to perform analysis on the uncertainty reduction for both the filter \cite{chen2014information} and smoother estimates, which identify practical information barriers. Propositions \ref{prop:asym_RE(FF)} and \ref{prop:asym_RE(OW)} outline the key findings from asymptotic analysis of the flow field and OW parameter smoother estimate via the relative entropy. The detailed derivations of which can be found in the Appendix Section \ref{Asec:deriv_asym_RE(FF)_RE(OW)}.

\begin{prop}[Asymptotic behavior of the uncertainty reduction in the flow field]\label{prop:asym_RE(FF)}

    Consider the Lagrangian data assimilation framework in Proposition \ref{prop:Asym_var_FF}. By substituting the asymptotic flow field smoother variance \eqref{eq:assym_smoother_var} into the relative entropy formula \eqref{eq:reletive_entropy_gaus}, the signal component converges to a constant while the dispersion component converges to a function that behaves like $\ln(L)$. Specifically, the uncertainty reduction in the variance of each velocity component in physical space has the form,
\begin{eqnarray}\label{eq:asym_disp_FF}
        \text{Dispersion}(\mathbf{u}_{\text{asym}})  &\sim&  C_1  +\frac{1}{4} \ln{\left(L\right)},
    \end{eqnarray}
where $C_1$ is a constant.

\end{prop}

The asymptotic behavior of the dispersion is approximated from the dominating behavior of the posterior variance, which is $\frac{1}{\sqrt{L}}$. This leads to the decay of the first term in \eqref{eq:reletive_entropy_gaus} while the last term grows as the negative power can be pulled out to the coefficient of the logarithm. The $\frac{1}{4}$ factor multiplying $\ln(L)$ is produced by the $\frac{1}{2}$ factor in the relative entropy formula \eqref{eq:reletive_entropy_gaus} multiplied by the $-\frac{1}{2}$ power on $L$, while the constant $C_1$ is determined by the other subdominant components of the relative entropy formula. It can be seen in Proposition \ref{prop:asym_RE(OW)} that a similar structure arises in the asymptotic behavior of the OW dispersion.

\begin{prop}[Asymptotic behavior of the uncertainty reduction in the OW parameter]\label{prop:asym_RE(OW)}

    Consider the same Lagrangian data assimilation framework in Proposition \ref{prop:Asym_var_FF}. By substituting the asymptotic OW smoother variance \eqref{eq:asym_var(OW)_general} into the relative entropy formula \eqref{eq:reletive_entropy_gaus}, the signal part converges to a constant while the dispersion component converges to a function of $\ln(L)$. Specifically, the OW parameter uncertainty reduction in the variance converges to a function of the form,
\begin{eqnarray}\label{eq:asym_disp_OW}
        \text{Dispersion}(\mbox{OW}(\mathbf{u}_{\text{asym}})) &\sim & C_2  +\frac{1}{4} \ln{\left( L \right) },
    \end{eqnarray}
    where $C_2$ is a constant.
\end{prop}

In the asymptotic limit, the dominating components of the OW variance are the mean-fluctuation terms as they scale with $\frac{1}{\sqrt{L}}$, which leads to the same factor in front of $\ln(L)$ as for the flow field. However, the constant $C_2$ will be distinct from $C_1$ as the remaining components of the dispersion are different from those of the flow field. The logarithmic form arising in both the flow field and OW parameter dispersion reveals a practical information barrier. For large numbers of tracers, each additional tracer provides less information. In fact, to reduce a fixed amount of uncertainty, an exponential increase in the number of tracers is needed.

\begin{figure}[h!]
    \centering
    \includegraphics[width=\linewidth]{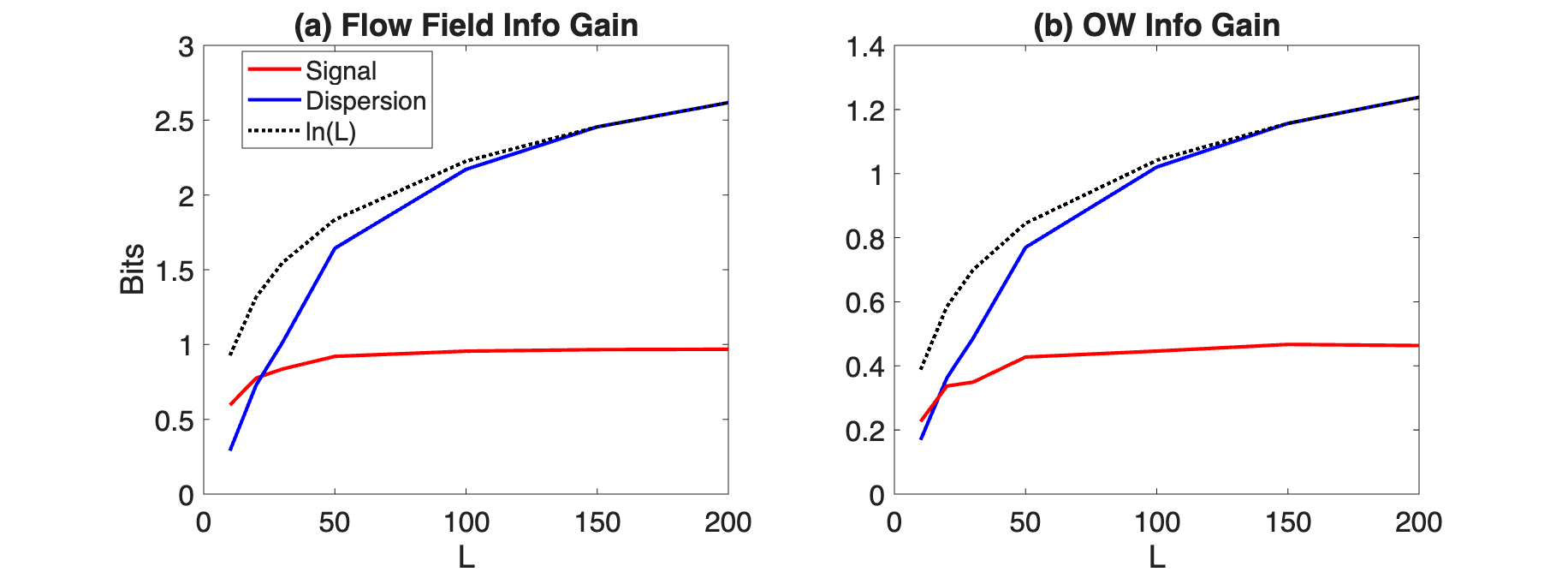}
    \caption{Average information gain from the prior distribution to the posterior distribution in the flow field (Panel (a)) and the OW parameter (Panel (b)). The red curve shows the signal, i.e., the uncertainty reduction in the mean from prior to posterior, whereas the blue curve is the dispersion, i.e., the uncertainty reduction in the variance. The black dashed curve shows a $\ln(L)$ curve for reference to evaluate the asymptotic behavior of the dispersion of the flow field and the OW parameter. The relative entropy is computed in physical space at each individual grid point and averaged across space and time.}
    \label{fig:RE_ave_curves}
\end{figure}

Figure \ref{fig:RE_ave_curves} shows the relative entropy of the flow field (Panel (a)) and the OW parameter (Panel (b)), broken into their signal (red) and dispersion (blue) components from the exact data assimilation results using the dynamical regime presented in Section \ref{sec:dyn_regime}. The black dotted line marks a function matching the asymptotic dispersion structure \eqref{eq:asym_disp_FF} and \eqref{eq:asym_disp_OW} for the flow field and OW parameter. The flow field and OW uncertainty reduction exhibit the same patterns: the uncertainty reduction in the mean (i.e. the signal) converges to a constant, while the uncertainty reduction in the covariance (i.e. the dispersion) goes like $\ln(L)$, aligning with the theoretical conclusions in propositions \ref{prop:asym_RE(FF)} and \ref{prop:asym_RE(OW)}.

\subsection{Uncertainty scaling with noise coefficient}

The asymptotic variance approximations \eqref{eq:assym_smoother_var}--\eqref{eq:asym_var(OW)_general} also explicitly relate the uncertainty and the model noise amplitude, enabling the framework to identify the dominating behavior of the prior and posterior estimates as the forecast system becomes increasingly turbulent. Let us use the noise coefficient $\sigma_{\mathbf{k}}$ in Equation \eqref{eq:dvk} as a baseline of the uncertainty in the flow field. As we increase the uncertainty in the system, the flow field and the OW parameter variance react differently. To provide concise results, let us assume all $\sigma_\mathbf{k}$ have the same value in the analysis here, which is consistent with the numerical setup. Proposition \ref{prop:noise_scaling} summarizes these theoretical differences.

\begin{prop}[Uncertainty in the flow field and the OW parameter as a function of $\sigma_{\mathbf{k}}$]\label{prop:noise_scaling}

    Consider the Lagrangian data assimilation scheme from Proposition \ref{prop:Asym_var_FF}, and assume $L\gg1$. Let $\sigma_\mathbf{k}$ vary with $\sigma_\mathbf{k}\ll L$. The flow field prior and posterior variances have clearly defined relations with $\sigma_\mathbf{k}$ \eqref{eq:FF_prior_var} and \eqref{eq:assym_smoother_var}. Furthermore, these relations directly translate to the fluctuation-fluctuation terms of the OW parameter variance \eqref{eq:Var(OW)_Fourier_general}, as they are solely quadratic functions of the flow field variance. On the other hand, the mean-fluctuation terms are not only linear in the flow field variance, but also incorporate additional dynamics through a quadratic function of the mean component. The dominating terms differ between the prior and posterior estimates, exhibiting distinct behaviors as $\sigma_\mathbf{k}$ varies.

    \begin{enumerate}%[label=(\roman*)]
        \item [(i)] \textbf{Prior Distribution}:
        \begin{itemize}
            \item $\mbox{Var}(\mathbf{u_{\text{prior}}}) \propto \sigma_{\mathbf{k}}^2$: The prior flow field variance is $\frac{\sigma_{\mathbf{k}}^2}{2d_\mathbf{k}}$, as defined by the stochastic flow field model \eqref{eq:dvk}.
            \item $\mbox{Var}(\mbox{OW}(\mathbf{u_{\text{prior}}})) \propto \sigma_{\mathbf{k}}^4 + \frac{f_\mathbf{k}^2}{d_\mathbf{k}^2}\sigma_{\mathbf{k}}^2$: The fluctuation-fluctuation and mean-fluctuation components of the prior OW variance will be on the same order and will both significantly contribute to the overall behavior of the OW variance. Thus, as the fluctuation-fluctuation terms form a quadratic function of the flow field variance and the mean-fluctuation terms a linear, the OW parameter variance will act as the sum of a quartic and quadratic function of $\sigma_\mathbf{k}$. Note that when the external forcings are zero, as is the case of the dynamical regime presented in Section \ref{sec:dyn_regime}, the quadratic term will disappear.
        \end{itemize}

        \item [(ii)] \textbf{Posterior Distribution}:
        \begin{itemize}
            \item $\mbox{Var}(\mathbf{u_{\text{post}}}) \propto \sigma_{\mathbf{k}}$: The posterior flow field variance will follow Equation \eqref{eq:assym_smoother_var}, which acts as a linear function of $\sigma_\mathbf{k}$ for $L\gg 1$.
            \item $\mbox{Var}(\mbox{OW}(\mathbf{u_{\text{post}}})) \propto \sigma_{\mathbf{k}}^3 + \left(\frac{f_{\mathbf{k}}^2}{d_{\mathbf{k}}^2} + \frac{f_{\mathbf{k}} f_{\mathbf{k'}}}{d_{\mathbf{k}}d_{\mathbf{k'}}}\right)\sigma_\mathbf{k}$: The mean-fluctuation terms of the OW parameter variance will dominate the OW variance as $L \gg 0$. The mean components converge asymptotically to their true solutions, which are produced by normal random variables with mean $\frac{f_{\mathbf{k}}}{d_{\mathbf{k}}}$, and standard deviation $\frac{\sigma_{\mathbf{k}}^2}{2d_\mathbf{k}}$, as defined by the stochastic flow field model \eqref{eq:dvk}. Thus the average value of $\overline{\hat{u}_\mathbf{k}} \hspace{0.2cm}\overline{\hat{u}_{-\mathbf{k}'}}$ across time will be $\frac{f_{\mathbf{k}} f_{\mathbf{k'}}}{d_{\mathbf{k}}d_{\mathbf{k'}}}$ when $\mathbf{k} \neq \mathbf{k}'$ and $\frac{f_{\mathbf{k}}^2}{d_{\mathbf{k}}^2} + \frac{\sigma_{\mathbf{k}}^2}{2d_\mathbf{k}}$ when  $\mathbf{k} = \mathbf{k}'$. The fluctuation component varies linearly in the flow field variance. Thus, the posterior OW variance as a whole will vary as the sum of a cubic and linear function of $\sigma_\mathbf{k}$. Note that under the zero forcing regime, the linear behavior will drop out.
        \end{itemize}
    \end{enumerate}
\end{prop}

The behavior of the uncertainty as a function of the model noise amplitude differs between the flow field and the OW parameter. In general, the OW parameter uncertainty acts as a higher power function of $\sigma_\mathbf{k}$ compared to the flow field. Thus, we have that the uncertainty of the OW parameter grows more rapidly than the flow field as the model becomes more turbulent. Additionally, the posterior variance of both the flow field and the OW parameter eliminates a factor of $\sigma_\mathbf{k}$ from the prior estimate, making the posterior estimate more controlled under increasingly turbulent model behavior.

\section{Connection between spatial patterns of the OW mean and variance}\label{sec:spatial_patterns}

In addition to uncovering key nonlinear relationships between the flow field and OW parameter statistics, as well as the asymptotic behavior of uncertainty, when applied to the OW parameter, the framework also reveals connections between the spatial patterns of the OW statistics. Specifically, the shared mean-fluctuation terms in the general and isotropic variance equations \eqref{eq:var(OW)_general_phy}  and \eqref{eq:var_phy_iso} are closely tied to the expectation of the OW parameter: the local minima of the expectation are associated with the local maxima of the variance. To get a clear picture of this connection, the isotropic case given in \eqref{eq:var_phy_iso} is primarily explored. However, the findings extend to the same terms in the general case \eqref{eq:var(OW)_general_phy}, while the terms not clearly tied to the expectation of the OW parameter may incorporate additional noise.

\begin{figure}[h!]
\centering
    \includegraphics[width=\linewidth]{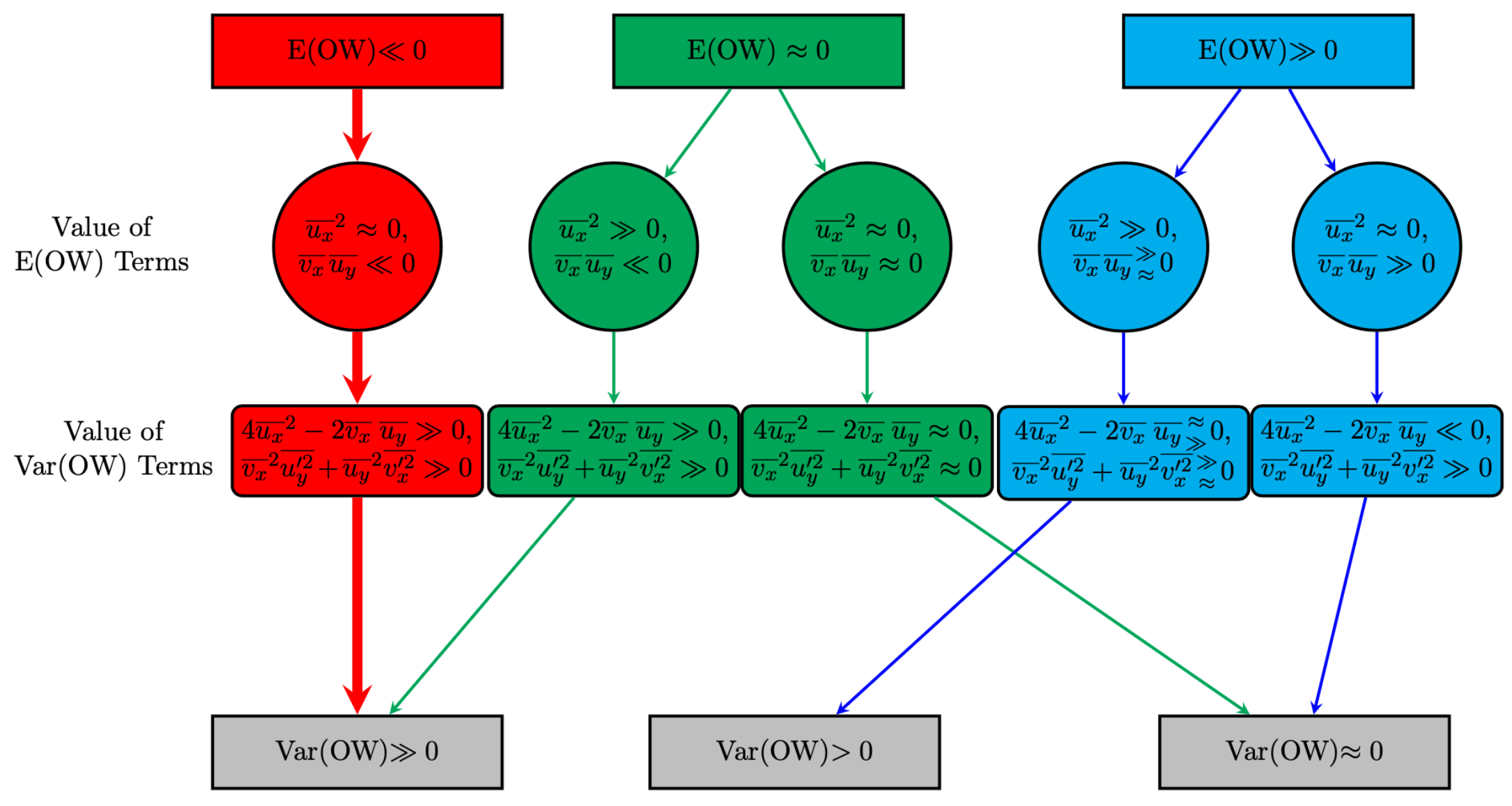}
\caption{Flow chart summarizing the link between spatial patterns in E(OW) and Var(OW). The top row of nodes shows cases of the overall value of E(OW), including large negative E(OW), small E(OW), and large positive E(OW). The second row of nodes describes the possible cases of the individual terms of E(OW). The third row shows the corresponding cases of the second-order fluctuation terms in Var(OW). Finally, the fourth row gives the resulting value of Var(OW) relative to other spatial points, including large variance resulting from both sets of second-order fluctuation terms being large, moderate variance resulting from only one set of terms being large, and small variance resulting from both sets of terms being small.}
    \label{fig:E(OW)_Var(OW)_flow_chart}
\end{figure}

Figure \ref{fig:E(OW)_Var(OW)_flow_chart} gives an overview of the connection between the expectation and the variance of the OW parameter, showing that only when the expectation of the OW parameter is large and negative is there a direct consequence in the variance (red). The top and bottom of the flow chart distinguish three key cases of the expectation and variance, respectively. The middle two rows indicate the associated cases of the individual terms. The expectation of the OW parameter \eqref{eq:E(OW)_incompres_PWA} has two terms: $\overline{u_x}^2$ and $\overline{v_x}\hspace{0.05cm} \overline{u_y}$. The variance \eqref{eq:var_phy_iso} has four mean-fluctuation terms. Here, these are paired as $\overline{u_x'^2}(4\overline{u_x}^2 - 2\overline{v_x}\hspace{0.1cm}\overline{u_y})$, and  $\overline{v_x}^2 \overline{u_y'^2}  + \overline{u_y}^2 \overline{v_x'^2}$, since the former closely resembles the structure of the expectation, and the latter is tied to the value of $\overline{v_x}\hspace{0.05cm} \overline{u_y}$ in the expectation. Notably, the only case of the expectation that directly ties to a single variance outcome is when the OW parameter is large and negative, which indicates strong vortical flow found at eddy centers. In such a case, the variance will be large as depicted in red in Figure \ref{fig:E(OW)_Var(OW)_flow_chart}. Thus, strong vortical flow is closely linked to high uncertainty. In contrast, when the expectation is small (green) or large and positive (blue), there are multiple possible outcomes for the OW variance, leading to a weaker relationship between the expectation and variance in these cases.

\begin{figure}[h!]
    \centering
    \includegraphics[width=\linewidth]{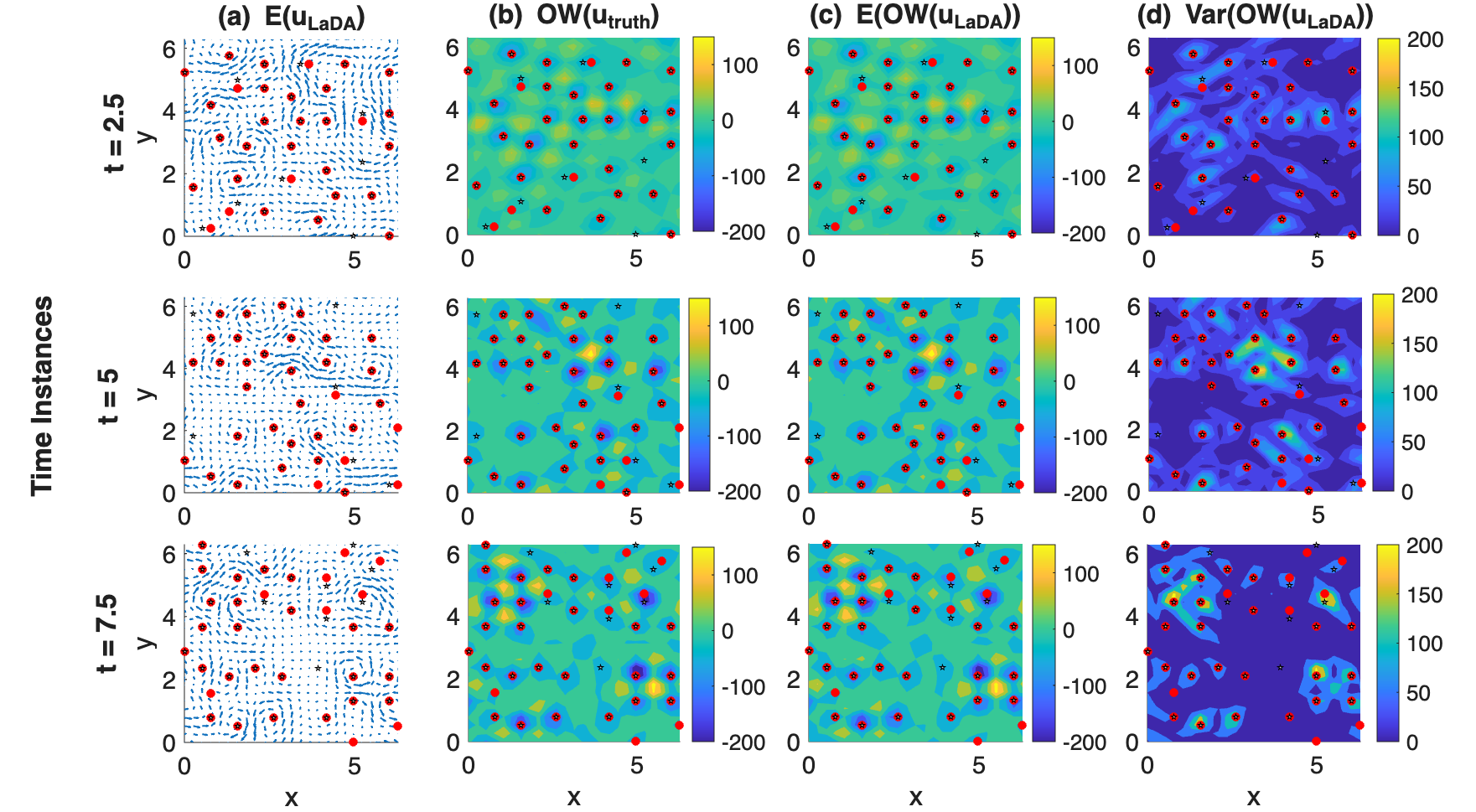}
    \caption{Behavior of the flow field and OW parameter for $L = 1000$ tracers. Each row depicts a snapshot of the flow field velocity smoother estimate (Panel (a)), the OW parameter evaluated on the true flow field used to generate the observations (Panel (b)), expected value of the OW parameter from the smoother estimate of the flow field using the Lagrangian data assimilation (LaDA) (Panel (c)), and variance of the OW parameter from the smoother estimate of the flow field (Panel (d)). The red dots identify the local minima of the E(OW($\mathbf{u}_{\text{LaDA}}$)) that are below the threshold value -0.2$\mbox{STD}(\mbox{OW}(\mathbf{u_{\text{LaDA}}}))$, and the light gray stars mark the local minima of OW($\mathbf{u}_{\text{truth}}$) below the threshold value -0.2$\mbox{STD}(\mbox{OW}(\mathbf{u_{\text{truth}}}))$. Here $\mbox{STD}(\mbox{OW}(\mathbf{u_{\text{truth}}}))$ and $\mbox{STD}(\mbox{OW}(\mathbf{u_{\text{LaDA}}}))$ are the spatial standard deviation. The flow field is approximated using the smoother estimate with 1000 tracers.}
    \label{fig:spatial_FF_EVOW_VarOW}
\end{figure}

This direct tie between the local minima of the expectation and the local maxima of the variance is apparent in spatial snapshots of the OW statistics. Figure \ref{fig:spatial_FF_EVOW_VarOW} examines the spatial patterns of the flow field and OW parameter statistics by comparing the smoother estimate with a large number of observations (1000 tracers) to the true OW values at three time points using the dynamical regime outlined in Section \ref{sec:dyn_regime}. For each snapshot, the local minima of the OW parameter from the smoother estimate and truth are shown as red dots and gray stars, respectively. These minima mark the areas of peak vortical flow, indicating eddy formation (Panel (a)). The expectation of the OW parameter (Panel (c)), and its local minima, align closely with the truth (Panel(b)), with very few spuriously identified or missed vortical flows, indicating that the estimated eddy criteria asymptotically converge to the truth. Notably, the local minima of the OW parameter expectation closely tie to the maxima of the variance as seen in Panel (d), numerically confirming the theoretical analysis outlined in Figure \ref{fig:E(OW)_Var(OW)_flow_chart}. Thus, eddy centers carry the most uncertainty in the flow field.

\begin{figure}[h!]
     \centering
     \includegraphics[width=\linewidth]{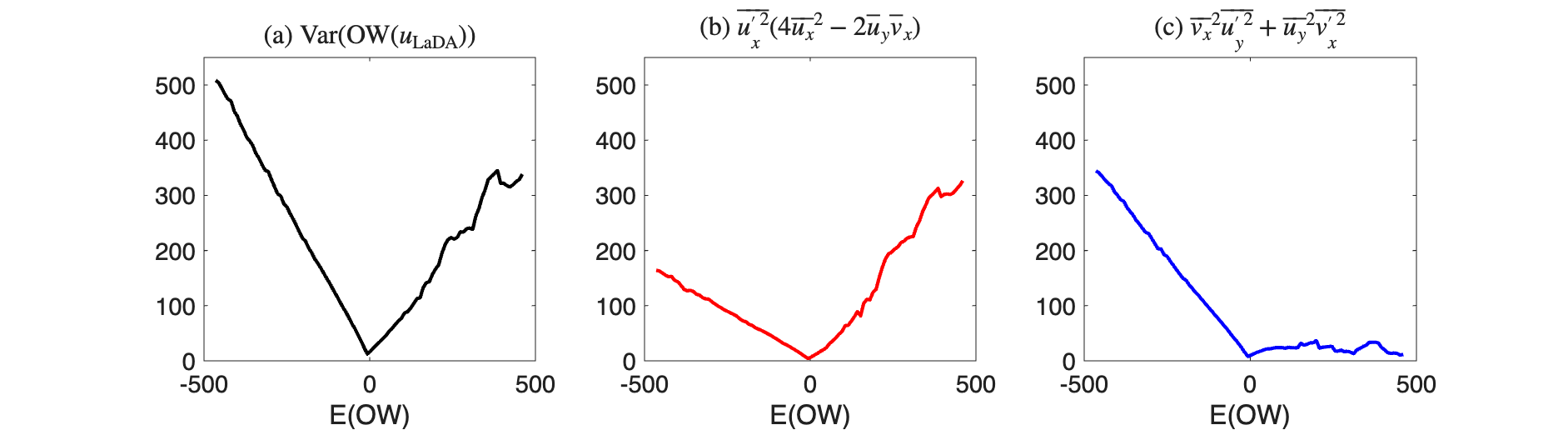}
     \caption{The total OW variance (Panel (a)) and the second order fluctuation terms in the variance formula (Panels (b) and (c)) plotted against the expected value of the OW parameter. The spatiotemporal points are partitioned according to the expected value of the OW parameter into 100 subsets (x axis). The values of the variance, and its second order fluctuation terms, are then averaged within each of these partitions (y-axis). }
     \label{fig:Var_v_E_OW}
 \end{figure}

 Figure \ref{fig:Var_v_E_OW} shows the average variance of the OW parameter \eqref{eq:var_phy_iso}, and the value of the two mean-fluctuation term pairs examined in Figure \ref{fig:E(OW)_Var(OW)_flow_chart}, as a function of the OW parameter expectation, to provide more insight into the overall spatial relationship between the OW statistics. The variance as a whole attains higher values when the expectation has a larger magnitude (Panel (a)). This is more pronounced for the large negative values of the expectation due to the direct tie to strong positive contributions from both term pairs, as seen in Panels (b) and (c), confirming the theoretical analysis in Figure \ref{fig:E(OW)_Var(OW)_flow_chart}. In contrast, while the terms depicted in Panel (b) show a faster rate of increase for positive values of the OW parameter due to the asymmetrical weighting, the positive expectations on average produce small amplitude values in the second term pair presented in Panel (c).

\begin{figure}[h!]
     \centering
     \hspace{-1cm}\includegraphics[width=\linewidth]{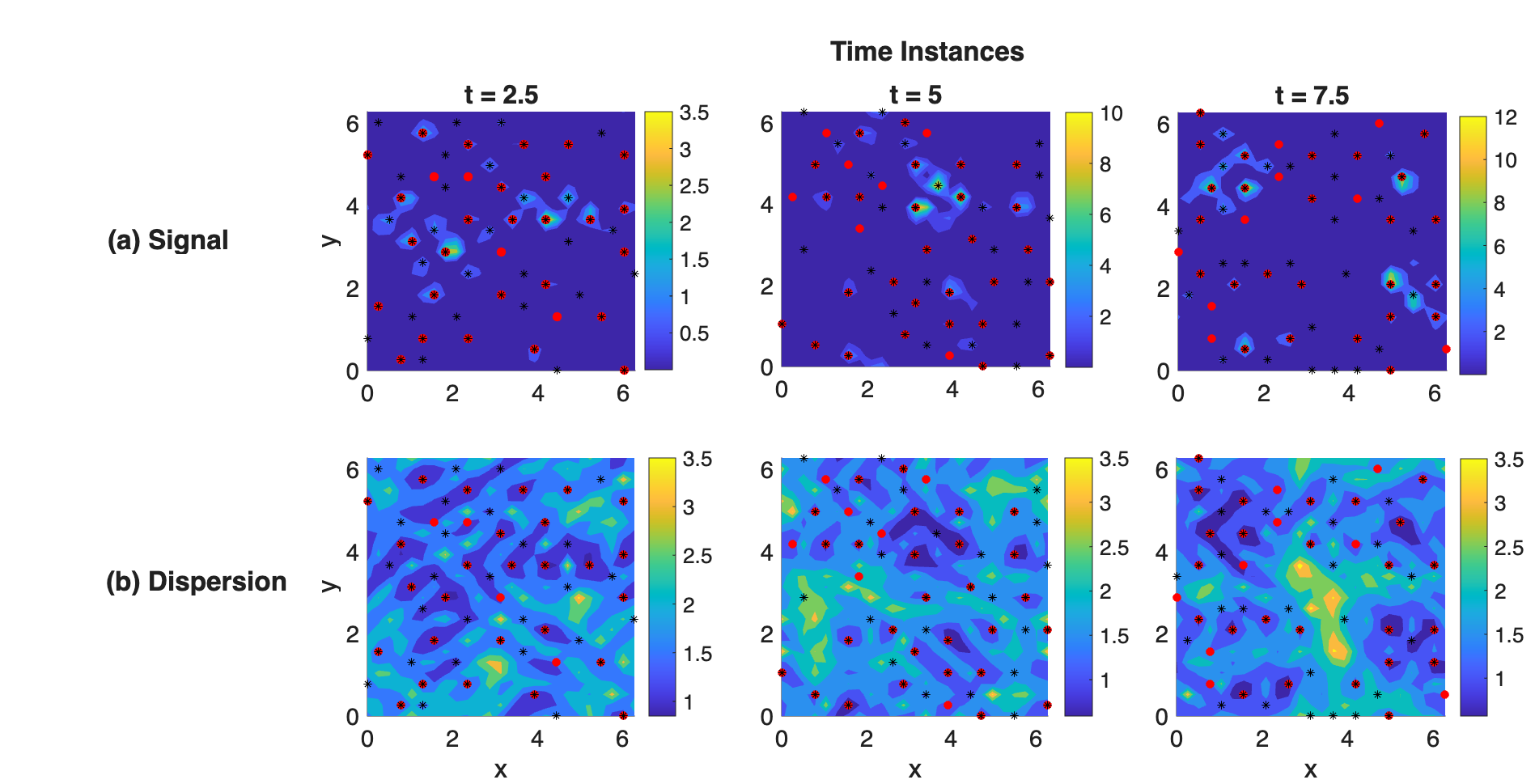}
     \caption{Snapshots of the signal (Panel (a)) and dispersion (Panel (b)) components of the relative entropy \eqref{eq:reletive_entropy_gaus}. The same time points as in Figure \ref{fig:spatial_FF_EVOW_VarOW} are shown for comparison. The red dots mark the same local minima of the expectation of the OW parameter as in Figure \ref{fig:spatial_FF_EVOW_VarOW}, and the black asterisks mark the local maxima of the signal component.}
     \label{fig:spatial_RE_1000t}
 \end{figure}

This connection between the OW parameter mean and variance carries into the uncertainty reduction, causing areas with large uncertainty reduction in the mean to correspond to limited uncertainty reduction in the variance, as shown in Figure \ref{fig:spatial_RE_1000t}. The local maxima of the uncertainty reduction in the mean (i.e., the signal Panel (a)) are marked by the black asterisks, while the local minima of the expectation of the OW parameter are marked by the red dots. Since the signal grows as the posterior mean deviates further from the prior, the local maxima of the uncertainty reduction in the mean correspond to the extrema of the OW parameter posterior expectation. This means that many of the local maxima of the signal will align with the local minima of the expectation, which are also areas with relatively large variance. The uncertainty reduction in the variance (i.e., dispersion Panel (b)) is small when the posterior variance is large, and thus the local maxima of the signal, as well as the local minima of the OW expectation, show little information gain in the variance. The least information gain occurs where the minima of the OW expectation and the maxima of the signal align, as these areas have the strongest vortical flow and thus the highest posterior uncertainty.

\begin{figure}[h!]
     \centering
     \includegraphics[width =\linewidth]{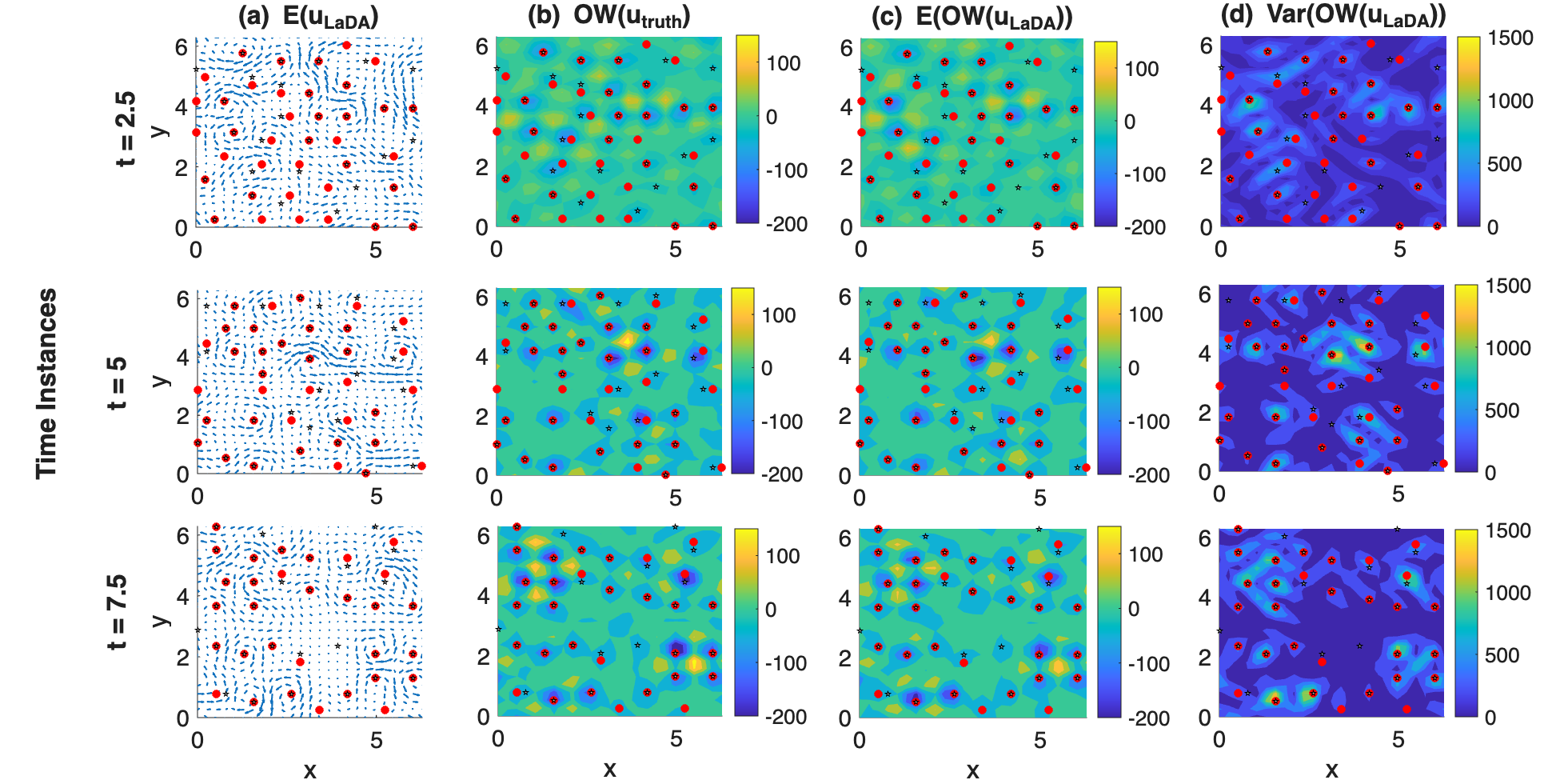}
     \caption{Spatial patterns of the approximate flow field velocity and OW parameter using 50 tracers to evaluate how the asymptotic conclusions extend to practical scenarios. The figure is structured similar to Figure \ref{fig:spatial_FF_EVOW_VarOW} but fewer tracers are used to approximate the flow field, and the exact smoother variance is used to compute the variance of the OW parameter.}
     \label{fig:spatial_FF_EVOW_VarOW_L50}
 \end{figure}

Similar to Figure \ref{fig:spatial_FF_EVOW_VarOW}, Figure \ref{fig:spatial_FF_EVOW_VarOW_L50} compares the spatial patterns of the flow field and OW parameter statistics to the true OW parameter, but for a limited number of tracers to demonstrate how the asymptotic conclusions translate to practical applications. For fewer tracers, the estimated OW parameter (Panel (c)) and its local minima do not perfectly align with the truth (Panel (b)). However, while less accurate than the 1000 tracer case, the majority of the same local minima are still identified close to their true location, with only a few weak vortical flows missed or spuriously identified (Panel (a)). Notably, the connection between the expectation and variance of the OW estimate persists with local minima of the expectation aligning with the local maxima of the variance (Panel (d)). Since this connection is produced by the underlying structure of the expectation and variance formulae and not an asymptotic estimate, it is sustained even for a small number of tracers.

\begin{figure}[h!]
    \centering
    \hspace{-1cm}\includegraphics[width=\linewidth]{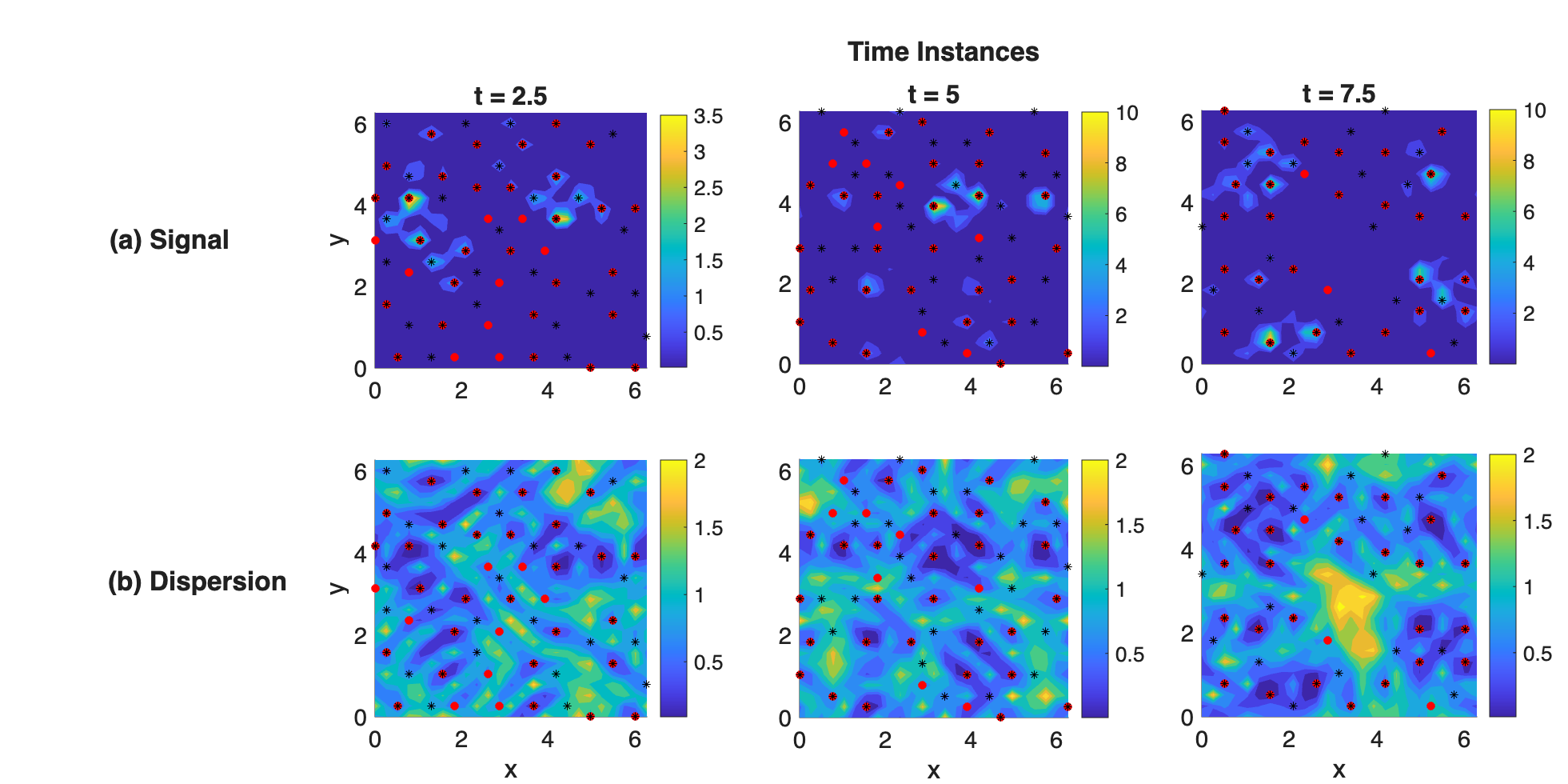}
    \caption{Snapshots of the signal (Panel (a)) and dispersion (Panel (b)) components of the relative entropy \eqref{eq:reletive_entropy_gaus}. Similar to Figure \ref{fig:spatial_RE_1000t} but for the flow field estimate given by only 50 tracers.}
    \label{fig:spatial_RE_50t}
\end{figure}

Since the tie between the expectation and variance of the OW parameter is preserved for a few tracers, the signal and dispersion spatial patterns maintain the same relationship as in the many tracer case. Figure \ref{fig:spatial_RE_50t} shows the signal and dispersion spatial patterns for the flow field estimate with 50 tracers. Again, many of the signal local maxima align with the local minima of the OW parameter expectation, as these are still areas where the posterior estimate deviates most from the prior (Panel (a)). Since these areas of vortical flow also have relatively high uncertainty even for a small number of tracers (shown in Figure \ref{fig:spatial_FF_EVOW_VarOW_L50}), the uncertainty reduction in the variance (Panel (b)) will be small in these areas, just as for the many tracer case presented in Figure \ref{fig:spatial_RE_1000t}.

\section{Discussion and Conclusion}\label{sec:conclusion}

The inherent nonlinearity in the eddy diagnostic presents complications, as it can amplify or suppress the uncertainty from the flow field in the resulting eddy criteria. In this paper, an analytically tractable mathematical and computational framework is developed to assess the uncertainty in eddy diagnostics. It balances analytic tractability with realistic turbulent dynamics by carefully incorporating both nonlinearity and stochastic parameterizations. The framework uses a simple stochastic flow field that effectively mimics key turbulent dynamics while facilitating theoretical study. Information theory is used to rigorously quantify uncertainty. A nonlinear, but analytically tractable, Lagrangian data assimilation scheme enables the study of uncertainty reduction via observations. The framework provides key insights by studying how the uncertainty interacts with the nonlinearity and how the uncertainty of the eddy diagnostic behaves as observations are incorporated. Not only does the framework provide a rigorous examination of the uncertainty, but it is also highly efficient, as it avoids the use of Monte Carlo simulations and instead relies on closed formulae. Notably, the asymptotic data assimilation solutions also facilitate the development of useful approximate schemes, which further accelerate the computation and allow the framework to be applied to more complicated problems in practice. See the Appendix Section \ref{Asec:Numerics} for details.

Applying the framework to the OW parameter reveals several key insights into the nonlinear uncertainty propagation:
\begin{itemize}
    \item The closed-form eddy diagnostic statistics reveal their explicit nonlinear relationship to the flow field statistics.
    \item Despite spatially homogeneous flow field uncertainty, the closed formulae show that the OW parameter variance can be inhomogeneous in space.
    \item The analytically tractable data assimilation scheme, alongside tools from information theory, allows the framework to identify a practical information barrier in the OW parameter uncertainty reduction.
    \item The closed OW parameter formulae also facilitate analysis of the dominating behavior of the uncertainty for a large number of observations.
    \item The closed-form eddy diagnostic statistics reveal that the local minima of the OW parameter expectation, corresponding to eddy centers, attain the largest uncertainties.
    \item Applying information theory reveals that areas with large uncertainty reduction in the OW mean correspond to areas of little uncertainty reduction in the OW variance.
\end{itemize}

The framework reveals several topics for future examination. Applying the framework to practical problems will provide several key insights into the uncertainty propagation. The closed formulae for the OW parameter expectation and variance enable rigorous uncertainty quantification across space and time. Furthermore, they allow for asymptotic analysis of the OW relative entropy, which can be used to inform decisions on data collection and how to optimize resources. Additionally, the link between the OW expectation and variance spatial patterns has strong implications for eddy identification, as the locations of eddy centers are most affected by uncertainty, emphasizing the need for uncertainty quantification in the eddy statistics, which is not addressed here. In addition, while this paper focuses on the OW parameter, the framework developed here is flexible and can be applied to other eddy criteria by computing the associated analytic formulae for their mean and variance. Furthermore, the framework can also be used to compare eddy criteria based on their sensitivity to the flow field uncertainty. Specifically, future work could address questions on the differences between uncertainty propagation through Eulerian and Lagrangian approaches.

\section*{Acknowledgement} 
N.C. is grateful to acknowledge the support of the Office of Naval Research (ONR) N00014-24-1-2244 and ONR MURI N00014-19-1-2421. C.M. is supported under the first grant above. S.W. acknowledges the financial support provided by the EPSRC Grant No. EP/P021123/1 and the support of the William R. Davis '68 Chair in the Department of Mathematics at the United States Naval Academy.

\section{Appendix}

\subsection{Derivations of the expectation and variance of the OW parameter}\label{Asec:deriv_E(OW)_Var(OW)}

Consider the two-dimensional incompressible random velocity field $\mathbf{u}(t,\mathbf{x}) = [u,v]^\mathtt{T}$, satisfying the spectral decomposition \eqref{eq:u_fourier}, with the mean-fluctuation decomposition \eqref{eq:mean_fluc_decomp}. Then, the OW parameter \eqref{eq:OW_incompressible} can be re-written as,
\begin{equation}\label{eq:MF_decomp_OW}
    \mbox{OW}(\mathbf{u}) = 4 (\bar{u}_x^2 + 2\bar{u}_xu_x' + u_x'^2 + \bar{v}_x\bar{u}_y + \bar{v}_xu_y' + v_x'\bar{u}_y + v_x'u_y').
\end{equation}
Taking the expectation gives
    $$\mbox{E(OW}(\mathbf{u})) = 4 (\overline{u_x}^2 + \overline{u_x'^2} + \bar{v}_x\bar{u}_y + \overline{v_x'u_y'}),$$
since $\mbox{E}(u') = \mbox{E}(v') = 0$. Furthermore, the Fourier decomposition reveals,
\begin{equation}\label{eq:Fourier_fluc_cancel}
    \overline{u_y'v_x'}  = - \overline{u_x'^2} = \sum_{\mathbf{k} \in \mathcal{K}, \mathbf{k \neq 0}} \frac{-k_1^2k_2^2\overline{\hat{u}_{\mathbf{k}}'(t)^2} }{k_1^2 +k_2^2}.
\end{equation}
Thus, the expectation can further be simplified to,
$$\mbox{E(OW}(\mathbf{u})) = 4 (\overline{u_x}^2  + \bar{v}_x\bar{u}_y ),$$
as shown in Proposition \ref{prop:E(OW)}.

To compute the variance, let us additionally assume a Gaussian flow field. Evaluating the variance of \eqref{eq:MF_decomp_OW} gives
$$ \mbox{Var}[\mbox{OW}(\mathbf{u})] = \mbox{E}[[4(\bar{u}_x^2 + 2\bar{u}_xu_x' + u_x'^2 + \bar{v}_x\bar{u}_y + \bar{v}_xu_y' + v_x'\bar{u}_y + v_x'u_y') - 4 (\bar{u}_x^2 + \overline{u_x'^2} + \bar{v}_x\bar{u}_y + \overline{v_x'u_y'})]^2]. $$
Then, by expanding, taking the expectation, and applying the property \eqref{eq:Fourier_fluc_cancel}, we obtain,
\begin{eqnarray}
    \nonumber\mbox{Var}[\mbox{OW}(\mathbf{u})] &=& 16\bigg[\overline{u_x'^4} + 2 \overline{u_x'^2v_x'u_y'} + \overline{v_x'^2 u_y'^2}  + 4 \bar{u}_x \overline{u_x'^3} + 2 \bar{v}_x \overline{u_x'^2u_y'} + 2\bar{u}_y \overline{u_x'^2 v_x'} \\
    \nonumber &+&  4\bar{u}_x \overline{u_x'v_x'u_y'} + 2 \bar{v}_x \overline{v_x'u_y'^2} + 2\bar{u}_y \overline{v_x'^2u_y'} + 4 \bar{u}_x^2 \overline{u_x'^2}\\
    \nonumber&+& 4\bar{u}_x \bar{v}_x \overline{u_x'u_y'} + 4\bar{u}_x\bar{u}_y \overline{u_x'v_x'} + \bar{v}_x^2 \overline{u_y'^2} - 2 \bar{v}_x \bar{u}_y \overline{u_x'^2} + \bar{u}_y^2 \overline{v_x'^2} \bigg].
\end{eqnarray}
Above, the terms are all fourth order, but have different compositions of mean and fluctuation components. Consider the terms with three fluctuation components and one mean component. Their Fourier representations all have the underlying form,
\begin{equation}
    \sum_{\mathbf{k}}\sum_{\mathbf{k}'}\sum_{\mathbf{k}'''} cE[\hat{u}_{\mathbf{k}}'\hat{u}_{\mathbf{k}'}'^*\hat{u}_{\mathbf{k}''}'^*] e^{i\mathbf{(k+k'+ k'') \cdot x}},
\end{equation}
where $\cdot^*$ denotes the complex conjugate. Since the modes are independent from each other, except their complex conjugates, the only terms that are not guaranteed to be zero are those with $k = k' = k''$. The remaining terms each have one of the following forms, which are all necessarily zero under the Gaussian flow field assumption,
$$ \mbox{E}(\hat{u}_{\mathbf{k}}'^3) = \mbox{E}( \hat{u}_{\mathbf{k}}'^* \hat{u}_{\mathbf{k}}'^2) = \mbox{E}(\hat{u}_{\mathbf{k}}'^{*2}\hat{u}_{\mathbf{k}}') = 0.$$
This can easily be shown by breaking each fluctuation component $\hat{u}_k'$ into its real and imaginary parts, $A'$ and $B'$, which are real-valued Gaussian random variables with mean $0$. For example, applying this to $\mbox{E}(\hat{u}_k'^3)$ yields,
$$\mbox{E}(\hat{u}_{\mathbf{k}}'^3) = \mbox{E}((A'+B'i)^3) = \mbox{E}(A'^3) - \mbox{E}(A')\mbox{E}(B'^2) + 3i\mbox{E}(A'^2)\mbox{E}(B') - \mbox{E}(B'^3)i -2\mbox{E}(B'^2)\mbox{E}(A') = 0.$$
The proof is similar for $\mbox{E}( \hat{u}_{\mathbf{k}}'^* \hat{u}_{\mathbf{k}}'^2)$ and $\mbox{E}(\hat{u}_{\mathbf{k}}'^{*2}\hat{u}_{\mathbf{k}}')$. Thus, all of the third order fluctuation terms are necessarily zero, and the variance formula can further be simplified to,
\begin{eqnarray}
    \nonumber\text{Var}[\text{OW}(\mathbf{u})] &=& 16\bigg[\overline{u_x'^4} + 2 \overline{u_x'^2v_x'u_y'} + \overline{v_x'^2 u_y'^2}  \\
    \nonumber &+&   4 \bar{u}_x^2 \overline{u_x'^2} + 4\bar{u}_x \bar{v}_x \overline{u_x'u_y'} + 4\bar{u}_x\bar{u}_y \overline{u_x'v_x'} + \bar{v}_x^2 \overline{u_y'^2} - 2 \bar{v}_x \bar{u}_y \overline{u_x'^2} + \bar{u}_y^2 \overline{v_x'^2} \bigg].
\end{eqnarray}

The primary goal here is to produce closed formulae for the OW parameter statistics that are completely determined by the flow field statistics, which are the mean and variance under the Gaussian flow field condition. To do so, it is necessary to express each expectation of a fourth-order fluctuation as the product of two expectations of second-order fluctuations. For the first term $\overline{u_x'^4}$, this follows directly from the Gaussian flow field assumption. Specifically, the kurtosis must satisfy,
$$\mbox{Kurt}(u) = \frac{\overline{u'^4}}{(\text{Var}(u'))^2} = \frac{\overline{u'^4}}{\overline{u'^2}^2} = 3.$$
Thus, $\overline{u'^4} = 3 \overline{u'^2}^2$. The remaining two fourth order fluctuation terms, can be simplified using Isserlis' theorem, which states,
$$\mbox{E}(u_1 u_2 u_3 u_4) = \mbox{E}(u_1u_2)\mbox{E}(u_3 u_4) + \mbox{E}(u_1u_3)\mbox{E}(u_2 u_4) + \mbox{E}(u_1u_4)\mbox{E}(u_3 u_2).$$
For the second term $\overline{u_x'^2v_x'u_y'}$, this gives,
$$\overline{u_x'^2v_x'u_y'} = \overline{u_x'^2} \overline{v_x'u_y'} + 2 \overline{u_x'v_x'} \hspace{0.2cm}\overline{u_x'u_y'} = -\overline{u_x'^2}^2 + 2 \overline{u_x'v_x'} \hspace{0.2cm}\overline{u_x'u_y'},$$
with the last equality following from \eqref{eq:Fourier_fluc_cancel}. Furthermore, the third term $\overline{v_x'^2 u_y'^2}$ can be expressed as,
$$ \overline{v_x'^2 u_y'^2} = \overline{v_x'^2}\hspace{0.2cm}\overline{u_y'^2} + 2\overline{v_x'u_y'}^2 = \overline{v_x'^2}\hspace{0.2cm}\overline{u_y'^2} + 2\overline{u_x'^2}^2,$$
again, with the last equality following from \eqref{eq:Fourier_fluc_cancel}. The variance formulae then becomes,
\begin{eqnarray}
    \nonumber\text{Var}[\text{OW}(\mathbf{u})] &=& 16\bigg[3\overline{u_x'^2}^2   + \overline{v_x'^2}\hspace{0.2cm}\overline{u_y'^2} + 4 \overline{u_x'v_x'} \hspace{0.2cm}\overline{u'_x u'_y} \\
    \nonumber&+& 4 \overline{u_x}^2 \overline{u_x'^2} - 2 \bar{v_x} \bar{u_y} \overline{u_x'^2} + \overline{v_x}^2 \overline{u_y'^2}  + \overline{u_y}^2 \overline{v_x'^2} +  4\bar{u_x} \bar{v_x} \overline{u_x'u_y'} + 4\bar{u_x}\bar{u_y} \overline{u_x'v_x'} \bigg],
\end{eqnarray}
as presented in Proposition \ref{prop:Var(OW)_general}, which has two types of terms referred to here as fluctuation-fluctuation and mean-fluctuation terms. To reveal additional information about the underlying structure of these terms we also derive the Fourier representation. Both classes of terms include the expectation of the second order fluctuations which all have the form,
 \begin{equation*}
    \sum_{\mathbf{k}} \sum_{\mathbf{k}'} c \overline{\hat{u}_{\mathbf{k}}'\hat{u}_{\mathbf{k}'}'^*} e^{i\mathbf{(k-k') \cdot x}} = \sum_{\mathbf{k}} c \overline{\hat{u}_{\mathbf{k}}'^2},
\end{equation*}
since the Fourier modes are independent. On the other hand, each second order mean component has the form,
$$\sum_{\mathbf{k}} \sum_{\mathbf{k}'} cc' \overline{\hat{u}_{\mathbf{k}}} \hspace{0.2cm}\overline{\hat{u}_{-\mathbf{k}'}} e^{i (\mathbf{k-k'}) \cdot\mathbf{x}}.$$
Thus, the full variance formula can be equivalently expressed by the Fourier representation,
\begin{eqnarray}
    \nonumber\text{Var}[\text{OW}(\mathbf{u})] &=& \sum_{\mathbf{k}} \sum_{\mathbf{k}'} \overline{|\hat{u}_{\mathbf{k}}'|^2} \hspace{0.2cm}\overline{|\hat{u}_{\mathbf{k}'}'|^2} (3a_1a_1' + 4b_1b_1' + c_1c_1')\\
    \nonumber &+& \sum_{\mathbf{k}} \sum_{\mathbf{k}'} \sum_{\mathbf{k}''} \overline{\hat{u}_{\mathbf{k}}} \hspace{0.2cm}\overline{\hat{u}_{-\mathbf{k}'}}\hspace{0.2cm}\overline{|\hat{u}_{\mathbf{k}''}'|^2} (4d_1d_1'd_2'' - 2f_1f_1'f_2'' + g_1g_1'g_2''\\
    \nonumber & & \hspace{4.4cm}  + h_1h_1'h_2''+ 4j_1j_1'j_2'' + 4l_1l_1'l_2'') e^{i (\mathbf{k-k'}) \cdot\mathbf{x}}.
\end{eqnarray}
The coefficients are all functions of the wave numbers, where functions of $\mathbf{k}'$, and $\mathbf{k}''$ are denoted by a `` $'$ '' and `` $''$ '', respectively. The functional forms of each coefficient are given in Table \ref{tab:Fourier_Var_coeffs}.

\begin{table}[h!]
    \centering
    \begin{tabular}{ c c c}
    %\multicolumn{1}{c}{} & \multicolumn{2}{c}{\textbf{}} \\
    %\multirow{2}{*}{\rotatebox{90}{\textbf{Row Label}}}\\
        \hline
        $a_1(\mathbf{k}) =  \frac{k_1^2k_2^2}{k_1^2 + k_2^2}$  &
        $a_1'(\mathbf{k}') = a_1(\mathbf{k}')$ & \\ \hline
        $b_1(\mathbf{k}) =  \frac{k_1^4}{k_1^2 + k_2^2}$&
        $b_1'(\mathbf{k}') = \frac{k_2'^4}{k_1'^2 + k_2'^2}$ &  \\ \hline
        $c_1(\mathbf{k}) =  \frac{-k_1^3k_2}{k_1^2 + k_2^2}$ &
        $c_1'(\mathbf{k}') = \frac{k_1'k_2'^3}{k_1'^2 +k_2'^2}$ & \\ \hline
        $d_1(\mathbf{k}) = \frac{k_1k_2}{\sqrt{k_1^2+k_2^2}}$ & $d_1'(\mathbf{k}') = d_1(\mathbf{k}')$  & $d_2''(\mathbf{k}'') = a_1(\mathbf{k}'')$ \\ \hline $f_1(\mathbf{k}) = \frac{-k_1'^2}{\sqrt{k_1'^2 +k_2'^2}}$ & $f_1'(\mathbf{k}') = \frac{k_2'^2}{\sqrt{k_1'^2 + k_2'^2}}$ & $f_2''(\mathbf{k}'') = a_1(\mathbf{k''})$ \\ \hline
        $g_1(\mathbf{k}) = f_1(\mathbf{k})$  & $g_1'(\mathbf{k}') = f_1(\mathbf{k}')$ & $g_2''(\mathbf{k}'') = b_1'(\mathbf{k}'')$ \\ \hline
         $h_1(\mathbf{k}) = f_1'(\mathbf{k})$ & $h_1'(\mathbf{k'}) = f_1'(\mathbf{k}')$ & $h_2''(\mathbf{k}'') = b_1(\mathbf{k}'')$ \\ \hline
        $j_1(\mathbf{k}) = d_1(\mathbf{k})$ &
        $j_1'(\mathbf{k}') = f_1(\mathbf{k}')$ & $j_2''(\mathbf{k}'') = c_1'(\mathbf{k}'')$ \\ \hline
        $l_1(\mathbf{k}) = d_1(\mathbf{k})$ & $l_1'(\mathbf{k}') = f_1'(\mathbf{k}')$ & $l_2''(\mathbf{k}'') = c_1(\mathbf{k}'')$\\
        \hline

    \end{tabular}
    \caption{Coefficients for the Fourier space variance formulae \eqref{eq:Var(OW)_Fourier_general} and \eqref{eq:Var(OW)_Fourier_iso}. Each row corresponds to a different term in the physical space equations \eqref{eq:var(OW)_general_phy} and \eqref{eq:var_phy_iso} specifically in order of \eqref{eq:var(OW)_general_phy}. The first three rows correspond to the fluctuation-fluctuation terms and the last six correspond to the mean-fluctuation terms.}
    \label{tab:Fourier_Var_coeffs}
\end{table}

By assuming an isotropic flow field, for which the equal energy partition used here falls under, the variance can be further simplified. Under the isotropic assumption, the asymptotic variance for large $L$ \eqref{eq:assym_smoother_var} will be equal for modes of equal magnitude. Now consider the mode pairs, $[k_1,k_2]$, $[k_1, -k_2]$. For most of the terms, the coefficients of these mode pairs will be equivalent since the variance coefficients have even powers of $k_1$ and $k_2$. However, for $c_1$, $c_1'$, and thus also for $j_2$ and $l_2$, the mode pairs $[k_1,k_2]$, $[k_1, -k_2]$ cancel each other since,
\begin{equation}
    \frac{-k_1^3 k_2}{\sqrt{k_1^2 +k_2^2}} \overline{|\hat{u}_{\mathbf{k}}'|^2}  + \frac{-k_1^3 (-k_2)}{\sqrt{k_1^2 +k_2^2}} \overline{|\hat{u}_{\mathbf{k}}'|^2}  = \frac{k_1 k_2^3}{\sqrt{k_1^2 +k_2^2}} \overline{|\hat{u}_{\mathbf{k}}'|^2}  + \frac{k_1 (-k_2)^3}{\sqrt{k_1^2 +k_2^2}} \overline{|\hat{u}_{\mathbf{k}}'|^2}  = 0.
\end{equation}
Thus we arrive at the isotropic definitions given in Proposition \ref{prop:Var(OW)_iso}, that are identical to the general variance definition without the terms $c$, $j$, and $l$.

\subsection{Derivations of the asymptotic posterior variance of the flow field and OW parameter}\label{Asec:deriv_asym_Var(FF)_Var(OW)}

 The filter covariance matrix, under the Lagrangian data assimilation scheme presented in Section \ref{sec:LaDA_scheme}, has previously been shown to converge to a diagonal with entries \cite{chen2023uncertainty},
\begin{equation}\label{eq:r_n_filter}
    R_{f\mathbf{k}} = \frac{\sigma_\mathbf{k}^2}{d_\mathbf{k} + \sqrt{d_\mathbf{k} + L\sigma_x^{-2}\sigma_\mathbf{k}^2}}.
\end{equation}
A similar property can be shown for the smoother covariance matrix. For our system, the backward time evolution of the smoother covariance is given by the following,
\begin{eqnarray}
    \nonumber \frac{\overleftarrow{\text{ d}\mathbf{R}_s}}{\d t} &=& -(-\mathbf{\Gamma} + (\mathbf{\Sigma}_U \mathbf{\Sigma}_U^*) \mathbf{R}^{-1}) \mathbf{R}_{s} - \mathbf{R}_{s}(-\mathbf{\Gamma}^* + (\mathbf{\Sigma}_U\mathbf{\Sigma}_U^*) \mathbf{R}^{-1}) +\mathbf{\Sigma}_U\mathbf{\Sigma}_U^*.
\end{eqnarray}
Note that $\mathbf{\Gamma}$, $\mathbf{\Sigma}_U$, and $\mathbf{R}$ are diagonal. This means, the $(m,n)$ element of $\frac{\d\mathbf{R}_{s}}{\d t}$ for $n\neq m$ is given by
$$\frac{\overleftarrow{\d\mathbf{R}}_{s;m,n}}{\d t} = -\left[(-d_m + i \omega_m) + \frac{\sigma_m^2}{r_m}\right] \mathbf{R}_{s;m,n} - \left[(-d_m - i \omega_m) +\frac{\sigma_n^2}{r_n}\right] \mathbf{R}_{s;m,n}.$$
Thus, the steady state gives
$$0 = -\left[(-d_m + i \omega_m) + \frac{\sigma_m^2}{r_m} + (-d_m - i \omega_m) +\frac{\sigma_n^2}{r_n}\right] \mathbf{R}_{s;m,n}.$$
For this to hold for all parameter values, $\mathbf{R}_{s;m,n}$ is necessarily equal to 0. Meaning, the smoother covariance matrix must be diagonal, with entries $R_{s\mathbf{k}}$ governed by,
\begin{equation*}
    \frac{\d R_{s\mathbf{k}}}{\d t} = 2d_n R_{s\mathbf{k}} - 2\frac{\sigma_\mathbf{k}^2}{R_{f\mathbf{k}}} R_{s\mathbf{k}} + \sigma_n^2.
\end{equation*}
The solution to the steady state equation is,
$$R_{s\mathbf{k}} = \frac{\sigma_\mathbf{k}^2}{-2d_\mathbf{k} + 2 \sigma_\mathbf{k}^2 R_{f\mathbf{k}}^{-1}},$$
which can be expanded using \eqref{eq:r_n_filter} to obtain,
\begin{equation*}
    R_{s\mathbf{k}} = \frac{\sigma_\mathbf{k}^2 }{2\sqrt{d_\mathbf{k} +  L \sigma_x^{-2}\sigma_\mathbf{k}^2}},
\end{equation*}
as defined in Proposition \ref{prop:Asym_var_FF}. This asymptotic flow field variance approximation can be substituted in the OW variance formula \eqref{eq:Var(OW)_Fourier_general} to obtain its asymptotic approximation \eqref{eq:asym_var(OW)_general}.

\subsection{Derivations of the asymptotic relative entropy in the flow field and the OW parameter}\label{Asec:deriv_asym_RE(FF)_RE(OW)}

In this study, the relative entropy \eqref{eq:reletive_entropy_gaus} is evaluated at individual grid points in physical space, in order to utilize the expectation and variance formulae \eqref{eq:E(OW)_incompres_PWA} and \eqref{eq:Var(OW)_Fourier_general}, and have comparable values between the flow field and OW parameter. In this form, the flow field relative entropy in the two-dimensional physical space is defined as,
\begin{eqnarray}
    \nonumber \mathcal{P}(p(\mathbf{u}(t,\mathbf{x})| \mathbf{X}(s \subset [0,T])), p(\mathbf{u}(t,\mathbf{x}))) &= &\\
    \nonumber& & \hspace{-1.5cm}\frac{1}{2}[(\mathbf{m}_{\mathbf{u}}-\mathbf{m}_{\mathbf{u}}^{att})^T(\mathbf{R}_{\mathbf{u}}^{att})^{-1}(\mathbf{m}_{\mathbf{u}}-\mathbf{m}_{\mathbf{u}}^{att})]  \quad \quad \quad \quad \quad \quad \hspace{0.08cm} \ldots \text{ Signal}\\
    \nonumber& & \hspace{-1.5cm}+ \frac{1}{2}[tr(\mathbf{R}_{\mathbf{u}}(\mathbf{R}_{\mathbf{u}}^{att})^{-1}) - 2 - \ln{\det(\mathbf{R}_{\mathbf{u}}(\mathbf{R}_\mathbf{u}^{att})^{-1})}] \quad \ldots \text{ Dispersion}.
\end{eqnarray}
Here, $\mathbf{m}_{\mathbf{u}}^{att}$ and $\mathbf{m}_{\mathbf{u}}$ are the prior and posterior means of the flow field in physical space defined by,
\begin{equation}
    \mathbf{m}_{\mathbf{u}} = \sum_{\mathbf{k}\in \mathcal{K}} \frac{\mu_{s\mathbf{k}}(t)}{\sqrt{k_1^2+k_2^2}} \begin{bmatrix}
        -ik_2\\
        ik_1
    \end{bmatrix}, \quad \mathbf{m}_{\mathbf{u}}^{att} = \sum_{\mathbf{k}\in \mathcal{K}} \frac{f_{\mathbf{k}}}{d_{\mathbf{k}}\sqrt{k_1^2+k_2^2}} \begin{bmatrix}
        -ik_2\\
        ik_1
    \end{bmatrix}.
\end{equation}
Additionally, $\mathbf{R}_{\mathbf{u}}$ and $\mathbf{R}_{\mathbf{u}}^{att}$, are the $2 \times 2$ covariance matrices for the zonal and meridional velocities $u$ and $v$ from the smoother estimate and the prior distribution, respectively. For sufficiently large $L$ they are defined by,
\begin{equation}
    \mathbf{R}_{\mathbf{u}} = \sum_{\mathbf{k}\in \mathcal{K}} \frac{R_{s\mathbf{k}}}{k_1^2+k_2^2} \begin{bmatrix}
        k_2^2 & -k_1k_2\\
        -k_1k_2 & k_1^2
    \end{bmatrix}, \quad \mathbf{R}_{\mathbf{u}}^{att} = \sum_{\mathbf{k}\in \mathcal{K}} \frac{\sigma_{\mathbf{k}}^2}{2d_{\mathbf{k}}(k_1^2+k_2^2)} \begin{bmatrix}
        k_2^2 & -k_1k_2\\
        -k_1k_2 & k_1^2
    \end{bmatrix}.
\end{equation}

The OW parameter relative entropy is similarly defined as,
\begin{eqnarray}
    \nonumber \mathcal{P}(p(\mbox{OW})| \mathbf{X}(s \subset [0,T])), p(\mbox{OW}) &= &\\
    \nonumber& & \hspace{-1.5cm}\frac{1}{2}[(\mathbf{m}_{\mbox{ow}}-\mathbf{m}_{\mbox{ow}}^{att})^T(\mathbf{R}_{\mbox{ow}}^{att})^{-1}(\mathbf{m}_{\mbox{ow}}-\mathbf{m}_{\mbox{ow}}^{att})]  \quad \quad \quad \quad \quad \quad \hspace{0.08cm} \ldots \text{ Signal}\\
    \nonumber& & \hspace{-1.5cm}+ \frac{1}{2}[tr(\mathbf{R}_{\mbox{ow}}(\mathbf{R}_{\mbox{ow}}^{att})^{-1}) - 1 - \ln{\det(\mathbf{R}_{\mbox{ow}}(\mathbf{R}_{\mbox{ow}}^{att})^{-1})}] \quad \ldots \text{ Dispersion}.
\end{eqnarray}
Note that for conciseness $\mbox{OW}(\mathbf{u}(t,\mathbf{x}))$ is shortened to OW here. The prior statistics, $\mathbf{m}_{\mbox{ow}}^{att}$ and $\mathbf{R}_{\mbox{ow}}^{att}$, and the posterior statistics, $\mathbf{m}_{\mbox{ow}}$ and $\mathbf{R}_{\mbox{ow}}$, are computed by inputting the prior and posterior flow field statistics into the OW expectation and variance formulae \eqref{eq:E(OW)_incompres_PWA} and \eqref{eq:var(OW)_general_phy}.

For the relative entropy's signal component of both the flow field and the OW parameter, the asymptotic behavior derivation is simple. As the number of observations increases, the posterior mean $\mathbf{m}_{\mathbf{u}}$ will converge to the true velocity, which is constant for each time instant $t$. Since the prior mean and variance are both constant, the flow field signal component will converge to a constant as the number of observations increases. Further, since the flow field expectation converges to a constant, the expectation of the OW parameter \eqref{eq:E(OW)_incompres_PWA} will also converge to a constant as the number of observations grows. Consequently, the OW parameter signal component also converges to a constant.

For the flow field dispersion, since $R_{s\mathbf{k}}$ asymptotically behaves like $1/\sqrt{L}$, $ \mathbf{R}_{\mathbf{u}}$ will behave as some constant matrix multiplied by the factor $1/\sqrt{L}$. Thus, since $\mathbf{R}_{\mathbf{u}}^{att}$ is a constant matrix, the dispersion will behave asymptotically like,
\begin{eqnarray}
    \text{Dispersion} (\mathbf{u}_{\text{asym}}) &\sim& \frac{1}{2}\left[\frac{c_1}{\sqrt{L}} - 2 - \ln{(c_2 L^{-1/2})}\right]\\
    \nonumber&=& \frac{1}{2}\left[\frac{c_1}{\sqrt{L}} - 2 -c_3 +\frac{1}{2} \ln{(L)}\right]\\
    \nonumber & \sim & C_1 +\frac{1}{4} \ln{(L)},
\end{eqnarray}
as described in Proposition \ref{prop:asym_RE(FF)}.

Similarly, since the asymptotic OW variance converges to $1/\sqrt{L}$, the dispersion will follow,
\begin{eqnarray}
    \text{Dispersion}(\mbox{OW}\mathbf{u}_{\text{asym}})) &\sim& \frac{1}{2}\left[\frac{c_4}{\sqrt{L}} - 1 - \ln{\left(c_5L^{-1/2}\right)} \right]\\
    &\sim& C_2 +\frac{1}{4} \ln{\left(L \right)} ,
\end{eqnarray}
as presented in Proposition \ref{prop:asym_RE(OW)}.

\subsection{Practical application of the asymptotic approximation}\label{Asec:Numerics}

%Highlight the use of the asymptotic approximation of the smoother variance in in practice: Note that for fewer tracers the variance amplitude is underestimated by the asymptotic approximation in the flow field which is exaggerated by the non-linearity. However, the spatial patterns of the variance are still very similar to the exact LaDA results. So while it may not necessarily be directly used beyond theoretical analysis in this context, it still holds helpful information, and may be used for other applications of data assimilation since it is rigorously rather than empirically derived.

In addition to providing crucial insights into the behavior of the uncertainty for a large number of tracers, the asymptotic approximation of the posterior variance can also be used to accelerate data assimilation. Rather than numerically solving the filter and smoother covariance equations \eqref{eq:R_f_LDA} and \eqref{eq:R_s_LDA}, which requires expensive matrix operations especially for larger problems, their asymptotic approximations \eqref{eq:r_n_filter} and \eqref{eq:assym_smoother_var} can be used as a rigorous estimate of the posterior variance. This approximation also simplifies the computation of the filter and smoother means \eqref{eq:mu_f_LDA} and \eqref{eq:mu_s_LDA} due to the diagonal structure of the posterior covariance estimates. Particularly, inverting the filter covariance in the computation of the smoother mean \eqref{eq:mu_s_LDA} is much more efficient using this approximation as the inverse of a diagonal matrix only requires inverting its diagonal elements.

\begin{figure}[h!]
     \centering
     \includegraphics[width=\linewidth]{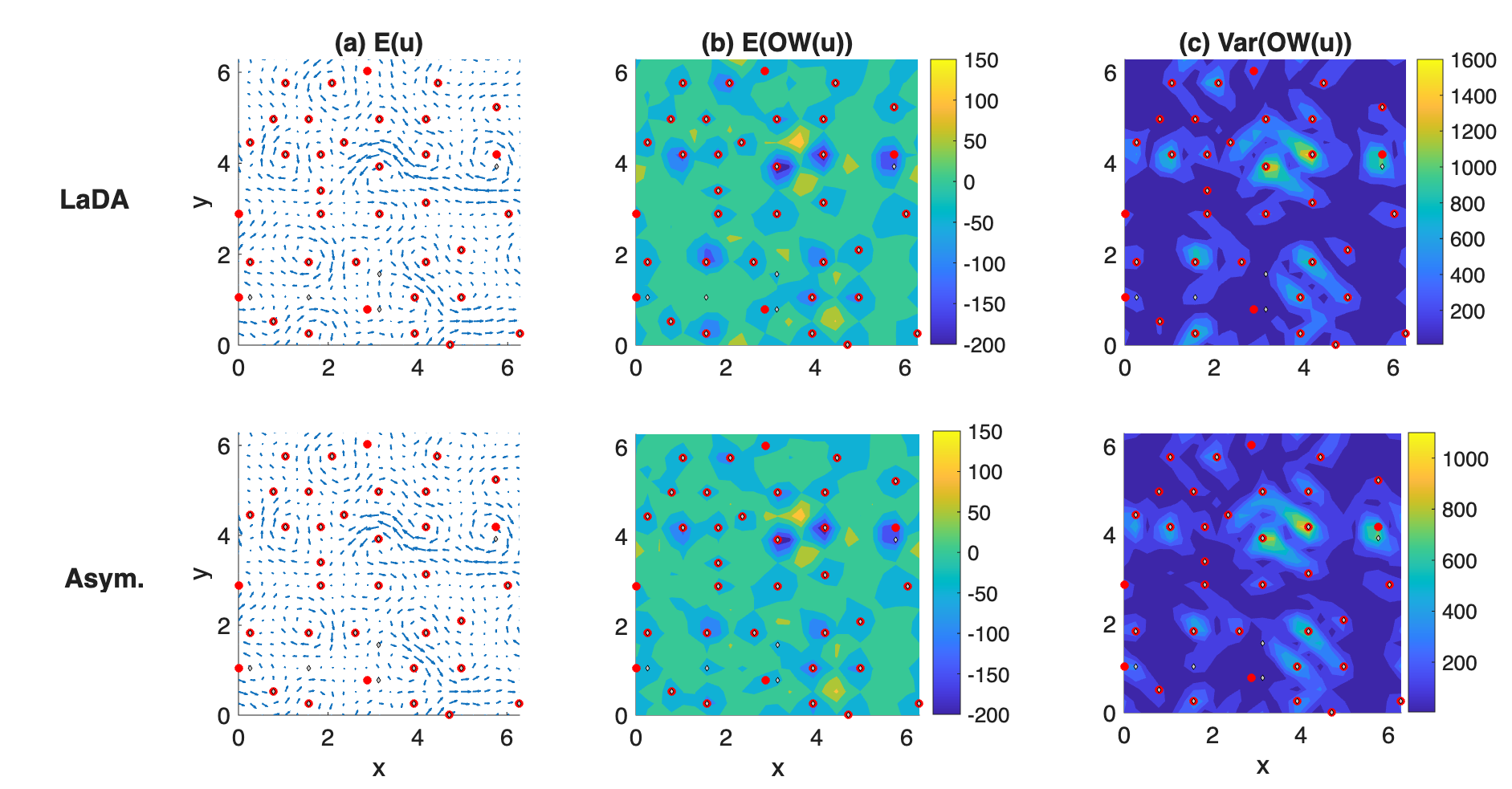}
     \caption{Snapshots of the flow field and OW parameter statistics at $t=5$ computed using the full Lagrangian data assimilation (LaDA) scheme \eqref{eq:mu_s_LDA}--\eqref{eq:R_s_LDA} (top row) and the asymptotic approximation (bottom row) both using 50 tracers. Panel (a) shows the expectation of the velocity field, (b) shows the expectation of the OW parameter, and (c) shows the variance of the OW parameter. The red dots mark the local minima $\mbox{E(OW}(u_{\text{LaDA}}))$ below the threshold value $-0.2\mbox{STD}(\mbox{OW}(\mathbf{u_{\text{LaDA}}}))$ same as in Figure \ref{fig:spatial_FF_EVOW_VarOW}, while the gray diamonds mark the local minima of $\mbox{E(OW}(u_{\text{asym}}))$ below the associated threshold$-0.2\mbox{STD}(\mbox{OW}(\mathbf{u_{\text{asym}}}))$.}
     \label{fig:asym_spatial}
 \end{figure}

While the application of this method for a large number of observations will be very close to the exact posterior estimate, for fewer tracers, which is often the case in practical applications, it is not necessarily a precise estimate. Nevertheless, the approximation can still provide a fairly accurate estimate that reveals crucial information about the exact data assimilation output. Figure \ref{fig:asym_spatial} shows a spatial snapshot of the flow field and the OW parameter statistics produced by the original Lagrangian data assimilation scheme \eqref{eq:mu_s_LDA}--\eqref{eq:R_s_LDA} compared to the asymptotic approximation using 50 tracers. The asymptotic approximation of the flow field closely matches the exact output of data assimilation (Panel (a)). Consequently, the spatial patterns of the OW parameter expectation (Panel (b)) and variance (Panel (c)) also closely resemble the exact posterior statistics. Many of the local minima of the approximated expectation (gray diamonds) align with the exact expectation (red dots) with very few missed, displaced, or spuriously identified. Furthermore, while the variance is underestimated by the approximation, the spatial patterns and their connection to the expectation of the OW parameter are maintained. So, while the asymptotic approximation may not necessarily be directly used beyond theoretical analysis to quantify uncertainty in the OW parameter, it still holds important information, and may be used for other applications of data assimilation as it is rigorously rather than empirically derived.

\newpage
\bibliographystyle{plain}
\bibliography{references}
\end{document}